\documentclass[12pt]{article}
\pdfoutput=1

\usepackage{amsmath,amssymb,amscd}
\usepackage{listings}
\usepackage{caption}
\usepackage{dsfont}
\usepackage{slashed}
\usepackage{color}
\usepackage{ulem}

\usepackage[pdftex]{graphicx}
\usepackage{epstopdf}
\usepackage{subfigure}
\usepackage{epsfig}
\usepackage{listings}
\usepackage{caption}
\usepackage{cite}

\usepackage{multirow}

\newcommand{\bea}{\begin{align}}
\newcommand{\eea}{\end{align}}
\newcommand{\beq}{\begin{equation}}
\newcommand{\eeq}{\end{equation}}
\newcommand{\nbea}{\begin{align*}}
\newcommand{\neea}{\end{align*}}
\newcommand{\nbeq}{\begin{equation*}}
\newcommand{\neeq}{\end{equation*}}
\newcommand{\bear}{\begin{eqnarray}}  
\newcommand{\eear}{\end{eqnarray}}  
\newcommand{\VL}{vector-like}  
\newcommand{\VLQ}{vector-like quark}

 \usepackage{multirow}
\usepackage{array}
\newcolumntype{M}[1]{>{\centering\arraybackslash}m{#1}}
\newcolumntype{N}{@{}m{0pt}@{}}

\numberwithin{equation}{section}

\begin{document}



\baselineskip=21pt
\rightline{{\fontsize{0.30cm}{5.5cm}\selectfont KCL-PH-TH/2015-56, LCTS/2015-46, CERN-PH-TH/2015-303}}
\rightline{{\fontsize{0.30cm}{5.5cm}\selectfont MCTP-15-33, CAVENDISH-HEP-15-14, DAMTP-2015-90}}
\vskip 0.3in

\begin{center}

{\large {\bf On the Interpretation of a \\
Possible $\sim 750$~GeV Particle Decaying into $\gamma \gamma$}}

\vskip 0.3in

 {\bf John~Ellis}$^{1,2}$,~
  {\bf Sebastian~A.~R.~Ellis}$^{3}$,~
   {\bf J\'er\'emie~Quevillon}$^{1}$,\\ ~
{\bf Ver\'onica Sanz}$^{4}$~
and {\bf Tevong~You}$^{5}$

\vskip 0.2in

{\small {\it

$^1${Theoretical Particle Physics and Cosmology Group, Physics Department, \\
King's College London, London WC2R 2LS, UK}\\
\vspace{0.25cm}
$^2${TH Division, Physics Department, CERN, CH-1211 Geneva 23, Switzerland}\\
\vspace{0.25cm}
$^3${Michigan Center for Theoretical Physics (MCTP), Department of Physics, University of Michigan, Ann Arbor, MI 48109, USA}\\
\vspace{0.25cm}
$^4${Department of Physics and Astronomy, University of Sussex, Brighton BN1 9QH, UK}\\
\vspace{0.25cm}
$^5${Cavendish Laboratory, University of Cambridge, J.J. Thomson Avenue, \\ Cambridge, CB3 0HE, UK;\\
\vspace{-0.25cm}
DAMTP, University of Cambridge, Wilberforce Road, Cambridge, CB3 0WA, UK}
}}

\vskip 0.25in

{\bf Abstract}

\end{center}

\baselineskip=18pt \noindent


{\small
We consider interpretations of the recent $\sim 3 \sigma$ reports by the CMS and ATLAS collaborations 
of a possible $X(\sim 750~{\rm GeV})$ state decaying into $\gamma \gamma$ final states.
We focus on the possibilities that this is a scalar or pseudoscalar electroweak isoscalar state produced by gluon-gluon fusion
mediated by loops of heavy fermions. We consider several models for these fermions, including a single
vector-like charge $2/3$ T quark, a doublet of vector-like quarks $(T, B)$, and a vector-like generation 
of quarks, with or without leptons that also contribute to the $X \to \gamma \gamma$ decay amplitude. 
We also consider the possibility that $X(750)$ is a dark matter mediator, with a neutral vector-like
dark matter particle. These scenarios are compatible with the present and prospective direct limits on vector-like fermions 
from LHC Runs 1 and 2, as well as indirect constraints from electroweak precision measurements, and we show that the
required Yukawa-like couplings between the $X$ particle and the heavy vector-like fermions are small enough to be 
perturbative so long as the $X$ particle has dominant decay modes into $gg$ and $\gamma \gamma$. 
The decays $X \to Z Z, Z \gamma$ and $W^+ W^-$ are interesting prospective signatures that may
help distinguish between different vector-like fermion scenarios. 
}


\vskip 0.2in

\leftline{ {\small December 2015}}

\newpage


\section{Introduction}

The CMS and ATLAS Collaborations have recently announced preliminary results
from the first $\sim 3$/fb of data from Run~2 of the LHC at 13~TeV, and both have
reported $\sim 3 \sigma$ enhancements in the inclusive $\gamma \gamma$ spectrum at $\sim 750$~GeV
that could be interpreted as decays of a possible massive particle $X$~\cite{Olsen,Kado}.
In the words of Laplace, {\it ``Plus un fait est extraordinaire, plus il a besoin d'\^etre appuy\'e de fortes preuves"}~\footnote{\it
``The more extraordinary a claim, the stronger the proof required to support it."},
so this evidence would need to be strengthened greatly before the existence of a new $X(750)$ state could
be regarded as established. Moreover, there are issues concerning the CMS and
ATLAS signals, e.g., the angular distributions of the $\gamma \gamma$ final states and the energy
dependence of the reported signal. Nevertheless, while maintaining our proper scepticism, we think it
worthwhile to explore possible interpretations of this possible new particle, and how they could be
probed experimentally, in the hope of either corroborating and elucidating the $X(750)$ signal or else
despatching it.

As in the case of the Higgs boson discovered in 2012~\cite{Eureka}, one may first ask what the spin
of the $X(750)$ particle could be. As in that case, the leading hypothesis would be spin zero, though
one should also consider spin two. The spin-two hypothesis would yield a
$\gamma \gamma$ angular distribution peaked in the beam directions~\cite{EHetal}. There
there is no significant evidence for this at the present time, but we consider the spin-two hypothesis
more exotic. Therefore,
we focus on spin-zero scenarios in the bulk of this paper, and on the corollary question whether the $X(750)$
could be scalar or pseudoscalar.

In either case the $\gamma \gamma$ decay mode reported would presumably arise from loop diagrams
with circulating fermions or bosons~\cite{anomaly}. Even if the $X(750)$ had couplings to the $t$ quark or $W^\pm$,
the form factors for their loops would be suppressed at large $\gamma \gamma$ invariant masses
and the dominant decays of the $X(750)$ would be to ${\bar t} t$ or $W^+ W^-$.
Hence the observation of the $\gamma \gamma$ decay mode
is {\it prima facie} indirect evidence for additional, heavier fermions and/or bosons whose masses are
$\gtrsim 750$~GeV. Having masses much greater than the electroweak symmetry-breaking scale,
any such fermions would presumably be vector-like, and much of this paper explores scenarios 
with massive vector-like quarks and/or leptons. Alternatively, the $\gamma \gamma$ decay could be
induced (partially) by loops of massive $W^\pm$ bosons, and we discuss the possibility that these
could correspond to the $\gtrsim 3 \sigma$ signal for a diboson resonance reported previously by
ATLAS and CMS.

Turning to possible production mechanisms for the $X(750)$, we recall that,
although each of CMS and ATLAS observe a $3-\sigma$ signal with $\sim 3$/fb at 13~TeV,
neither reported a signal with $\sim 20$/fb at 8~TeV~\cite{zilchATLAS,zilchCMS}, although there
is a small enhancement in the CMS data at $\sim 750$~GeV. The data
at different energies would be accommodated more easily if the $X(750)$ were produced via a mechanism with a steeper
energy dependence. From this point of view, and assuming that the $X(750)$ is
not produced in association with any other particle, gluon-gluon fusion would be
a more promising mechanism than ${\bar q} q$ annihilation
(though the energy-dependence does not favour greatly this
mechanism, and heavy ${\bar q} q$ annihilation would be preferred). Moreover, gluon-gluon fusion is favoured by historical
precedent (the Higgs boson) and by Occam's razor, since loops of heavy fermions could provide
this production mechanism as well as the $\gamma \gamma$ decay mode. Accordingly, in
later sections of this paper we concentrate on the possibility that gluon-gluon fusion is the dominant
production mechanism for the $X(750)$.

What fermions might generate the $gg \to X \to \gamma \gamma$ signal?
The chirality of the Standard Model (SM) under the electroweak SU(2)$_L \times$U(1)$_Y$ gauge symmetries 
requires a $\Delta I = 1/2$ Higgs boson to generate masses for elementary fermions, and electroweak precision tests 
exclude a fourth chiral generation of SM fermions at 7$\sigma$ \cite{Agashe:2014kda}.
Moreover, current bounds on the masses of new quarks from direct searches would require Yukawa couplings 
that is ${\cal O}(4)$ and hence unpalatably large. On the other hand, vector-like fermions $\chi$ could have 
gauge-invariant bilinear mass terms, $m_\chi \bar{\chi}\chi$, that are not tethered to the electroweak scale. However,
by the same token, such a bilinear mass term poses an additional hierarchy problem. 
Explaining how and why vector-like fermions masses could be near the electroweak scale 
is a rich topic of research which we will not go into here, though we cannot resist remarking that their lightness may provide
further motivation for supersymmetry (SUSY) or compositeness.

Setting aside this hierarchy problem, there is no known reason why vector-like fermions should not exist at or below the TeV scale. Indeed, 
they appear in many theories of beyond the Standard Model (BSM) physics, and are sometimes even necessary. For example, even the
minimal supersymmetric extension of the SM (the MSSM) contains vector-like fermions in the form of the Higgsinos, 
which are effectively a pair of vector-like lepton SU(2)$_L$ doublets~\footnote{However, loops of
MSSM sparticles could not explain the $X(750) \to \gamma \gamma$ signal.}.
In many string theories, such as D-brane theories \cite{Dijkstra:2004cc}
or heterotic string compactifications \cite{Lebedev:2006kn}, vector-like fermions occur quite frequently, often in complete vector-like families 
with SM-like charges. From a bottom-up perspective, vector-like families are often found in composite Higgs 
models~\cite{Contino:2006qr,Anastasiou:2009rv,Vignaroli:2012sf,DeSimone:2012fs,Delaunay:2013iia,Gillioz:2013pba}, 
little Higgs models~\cite{Han:2003wu,Carena:2006jx,Matsumoto:2008fq,Berger:2012ec}, scenarios with warped extra 
dimensions~\cite{Gopalakrishna:2013hua} and SUSY models beyond the 
MSSM~\cite{Kang:2007ib,Martin:2009bg,Graham:2009gy,Martin:2010dc,Moroi:2011aa,Martin:2012dg,Fischler:2013tva,Endo:2011xq,Endo:2012cc}. Recently, vector-like fermions have been considered in the context of the decay of a CP-odd scalar to vector bosons \cite{Gopalakrishna:2015wwa}. 

In this paper we take an agnostic attitude towards the possible origin and nature of vector-like fermions, and
consider the following representative scenarios, always assuming that the $X(750)$ is an SU(2) singlet: (i) $X$ is
coupled to an SU(2)-singlet vector-like top partner, (ii) $X$ is coupled to an SU(2)-doublet vector-like quark partner,
(iii) $X$ is coupled to a vector-like copy of a generation of SM quarks, i.e., one SU(2) doublet and two singlets, all with SM-like charge 
and hypercharge assignments, (iv) $X$ is coupled to a complete vector-like generation of SM-like quarks and leptons.
We estimate the required $X$ coupling as a function of the masses of the vector-like fermions in these models,
and we consider in each case their possible signatures, including indirect constraints from precision electroweak data,
flavour physics and dark matter relic density as well as direct LHC searches for the decays of heavy particles.

The outline of this paper is as follows. In Section 2 we present a general analysis of the production of a scalar $S$
or a pseudoscalar $P$ with a mass $\sim 750$~GeV via gluon fusion through loops of massive vector-like quarks, and its
subsequent $\gamma \gamma$ decay via analogous loops, including also the possibility of massive vector bosons. If a single
\VLQ\ were to contribute, we find that it would require quite a large $S/P$ coupling. However, this requirement would be relaxed if there
were more vector-like quarks, or if heavy bosons
also contributed to the $\gamma \gamma$ coupling. In Section 3 we introduce the four \VL\ fermion models we consider.
Section 4 we present some of the diboson decay signatures of these models, confronting them with the corresponding
experimental sensitivities, and Section 5 summarizes our conclusions. Finally, in an Appendix we give details of the models
in two-component notation for the \VL\ fermions.

\section{General Aspects of the $X \to \gamma \gamma$ Signal}

\begin{figure}
\centering
\includegraphics[scale=0.8]{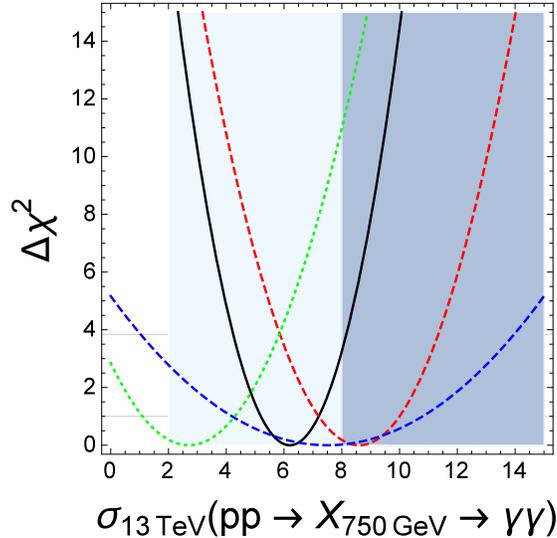}
\caption{\it The $\chi^2$ function for the 750 GeV resonance production cross-section times $\gamma\gamma$ branching ratio, in units of femtobarns, at 13 TeV for CMS (ATLAS) Run 2 results displayed in dashed blue (red) and for CMS Run 1 results in green dotted lines. The combination is shown in solid black with a best fit value and formal one-$\sigma$
range of $6.2 \pm 1.0$ fb. The 95\% CL exclusion from CMS Run 1 ranges from 2 to 8 fb corresponding to the shaded light and dark blue regions.}
\label{fig:chisquared}
\end{figure}

The best-fit cross-section for the signal at 13 TeV can be estimated by reconstructing the likelihood, assumed 
here to be essentially Gaussian, from the 95\% CL expected and observed limits as
was done for the Higgs boson in~\cite{Continoetal1202.3415}. We assume a resonance mass of 750 GeV and use the 95\% CL 
ranges from ATLAS and CMS at 13 TeV~\cite{Olsen,Kado} and CMS at 8 TeV~\cite{zilchCMS} (the ATLAS 8 TeV exclusions do not extend up to 750 GeV~\cite{zilchATLAS}). These are reported for narrow widths, which do not vary much at 750 GeV for widths below $\sim$ 10 GeV,
as shown in Fig.~9 of~\cite{zilchCMS}. The excess remains significant at both narrow and wide widths, 
with a slight preference from ATLAS for the latter but, given the limited information publicly available,
here we combine the best fits for the reported narrow width exclusions as an indicative cross-section range. 

Fig.~\ref{fig:chisquared} displays the resulting global $\chi^2$ function for the fit for the 750 GeV resonance production 
cross-section times $\gamma \gamma$ branching ratio at 13 TeV. The individual CMS (ATLAS) Run 2 results 
are shown as blue (red) dashed lines
while the CMS Run 1 result is shown as a green dotted line, where we have rescaled from 8 TeV to 13 TeV as described in
detail below. The combination is displayed as a solid black line, with the best-fit cross-section value and 68\% C.L.
range found using the method of~\cite{Continoetal1202.3415} to be $6.2 \pm 1.0$ fb~\footnote{The method 
of~\cite{Continoetal1202.3415} assumes a Gaussian approximation to reconstruct the likelihood which, as they note,
becomes accurate only when the number of events $N \gtrsim 10$. With the current limited data this estimate
deviates from an estimate based on Poisson statistics, but we use this method to give a rough indication of the signal cross-section 
region of interest should the signal grow with more statistics, recognising that the formal error it yields is probably an
underestimate.}.

The $X$ particle could be produced by a $q\bar q$ or a $gg$ intial state but, as already mentioned,
we assume here the gluon-initiated production mechanism, which is better able to accommodate the increase of 
the signal significance from LHC Run~1 at 8~TeV to LHC Run~2 at 13~TeV.

It is important to take into account the increase in the background as well as the energy dependence of the signal in estimating the relation between the observations at Run 2 and the exclusion limits by Run 1 searches. We can quantify the increase
in the signal significance via the double ratio
\bear
{\cal R}_{i} = \frac{(\sigma^{i}_{S}/\sqrt{\sigma_{B}})_{\text{13 TeV}}}{(\sigma^{i}_{S}/\sqrt{\sigma_{B}})_{\text{8 TeV}}} \, ,
\label{ratio}
\eear
where $i= $ $gg$, $q \bar q$, and $\sigma$ are the cross sections of signal ($S$) and 
background ($B$). If one rescales (\ref{ratio}) with the appropriate integrated luminosities ($\sim 20$/fb
for Run~1 and $\sim 3$/fb for Run~2) this ratio corresponds to the expected statistical increase in the number of
standard deviations from the 8-TeV run to the 13-TeV run.
We find that the increases for the two production mechanisms are
\bear
{\cal R}_{gg} \simeq 3 \, \textrm{, whereas } {\cal R}_{q \bar q} \simeq 1.7 \ .
\eear
These double ratios are largely insensitive to the mass of the resonance in a range of $M_{X}\simeq 700-800$ GeV,
and to the spin and CP properties of the resonance, e.g. $J^{CP} =$ $0^{+}$, $0^-$ and $2^+$. 
The spin of the resonance alters the kinematics, though, leading to a different $\gamma$ distribution in the rapidity bins. 

We evaluated the background events by simulating the main irreducible background ($pp\to \gamma\gamma$)
using  {\tt Madgraph}~\cite{MG5} at LO and performed a cut $|M_{\gamma\gamma}  - M_{X} | \leq 0.05 M_{X}$, 
as well as $|\eta_\gamma|<3$.  In principle, there are additional reducible backgrounds from 
$\gamma$ + jet and dijet events, but Fig.~2 of~\cite{zilchCMS}
indicates that these are small compared with the irreducible background for invariant masses $\sim 750$~GeV. 
We estimated  the NLO K-factor for a $gg$-initiated resonance by computing a heavy Higgs K-factor with 
MCFM~\cite{MCFM}. This K-factor is ${\cal O}(100\%)$, although its dependence roughly cancels out in the double ratio.

The cross-section excluded at the 95\% CL by the absence of a signal in the CMS Run 1 data~\cite{zilchCMS}
is approximately 0.5-2 fb for a spin-zero resonance with mass in the range of 700-800 GeV. 
This Run~1 limit can be translated into a 95\% CL upper limit on the allowed cross-section at 13 TeV
using the value of ${\cal R}$:
 \bear
 \sigma_{X} (\textrm{LHC13})  \lesssim \, 4.2 \,   \sigma_{X}(\textrm{LHC8}) \simeq (2-8) \textrm{ fb, }
 \eear
where we have used ${\cal R}\simeq 3$ and $\sigma_{B} (\textrm{LHC13})/ \sigma_{B} (\textrm{LHC8}) \simeq 2$.
The excluded cross-section from CMS Run 1 depends on the assumed total decay width, with typically stronger limits for narrower widths, but the uncertainty in the signal-to-background ratio does not allow a more meaningful extrapolation from 8 to 13 TeV of the limits, other than the broad range of 2-8 fb that we calculated here, which seems completely compatible with the
strengths of the signals reported by CMS and ATLAS. The 2 (8) fb exclusions by CMS Run 1 are shaded in light (dark) blue in Fig.~\ref{fig:chisquared}, and we see that the combined best-fit cross-section is within $\sim 2$ sigma of the weakest exclusion. More data will be needed to answer whether there is a statistically significant incompatibility between
the 8 and 13 TeV data that requires further explanation. 
 
\section{The $X$ Couplings to Vector Bosons}

In the following we focus on a spin-zero $X$ particle, considering two options for the CP properties, 
namely a scalar and a pseudoscalar state.  Possible UV origins of the scalar resonance are a 
dilaton~\cite{dilaton} from the breaking of conformal invariance, or equivalently a 
radion~\cite{radion} from an extra dimension.  A pseudoscalar particle could also have several origins, 
e.g., an axion-like particle from the breaking of a Peccei-Quinn symmetry~\cite{PQ}, 
or a pseudo-Goldstone boson from symmetry breaking in a composite Higgs model~\cite{CHGripaios}. 
One could also contemplate the possibility that the resonance at 750 GeV is part of an extended Higgs sector, 
such as a 2-Higgs-doublet model (2HDM) that might originate from supersymmetry. 
Alas, in a 2HDM the coupling to fermions and gauge bosons is constrained,
leading to a branching ratio to photons two orders of magnitude below what would be required to explain the 
signal. In this paper we consider a different approach, with new heavy fermions inducing the coupling of the 
resonance to gauge bosons.  

\begin{figure}[h!]
\centering
\includegraphics[scale=0.8]{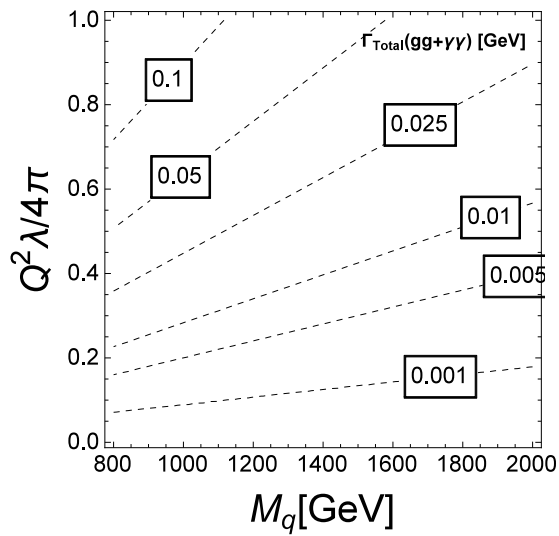}
\caption{\it Total decay width in GeV assuming the dominance of decays to gluon gluon and $\gamma \gamma$ final states
mediated by a single heavy vector-like quark of charge $Q$ and mass $M_q$. }
\label{fig:totaldecaycontours}
\end{figure}

Irrespective of the specific origin of the resonance, the couplings of a generic scalar $S$ and 
pseudoscalar $P$ to pairs of photons and gluons are described via dimension-five operators in an effective field theory (EFT):
\bear
{\cal L}_{eff} =&-& S\, \left(\frac{g_{S\gamma}}{4} \,F_{\mu\nu} F^{\mu\nu} + \frac{g_{Sg}}{4}  \,G_{\mu\nu} G^{\mu\nu}\right) \nonumber \\
&-&  P \, \left(\frac{g_{P\gamma}}{4} \,F_{\mu\nu} \tilde{F}^{\mu\nu} + \frac{g_{Pg}}{4}  \,G_{\mu\nu} \tilde{G}^{\mu\nu}\right) 
\label{eq:Leff}
\eear
Within the EFT, one can compute the partial widths of the $X$ to gluons and photons as
\begin{equation}
\Gamma_\text{EFT}(X \to g g) \;  = \; \frac{g_{X g}^2}{8 \pi} \, m_X^3 \, ,  \; \Gamma_\text{EFT}(X \to \gamma \gamma) \; = \; \frac{g_{X\gamma}^2}{64 \pi} \, m_X^3 \, ,
\label{eq:GammaEFTS}
\end{equation}
where $X=S$ or $P$. 
The total decay width is very small if we assume domination by these
decays into gluons and photons. For example, in Fig.~\ref{fig:totaldecaycontours} we display contours of widths 
including only decays into gluons and photons for a typical model with a heavy vector-like quark of charge $Q$ 
responsible for the loop-induced coupling, as a function of the mass of the quark and its coupling $\lambda$ to the scalar. 
Although ATLAS reports that its significance is largest for a width of 6\% of $m_X$~\cite{Kado}, 
the excess remains almost as significant for narrow widths. In the following we treat the decay width 
as a free parameter and plot the parameter space for both a narrow width as above and a wide width of 45 GeV. 

\begin{figure}
\centering
\includegraphics[scale=0.8]{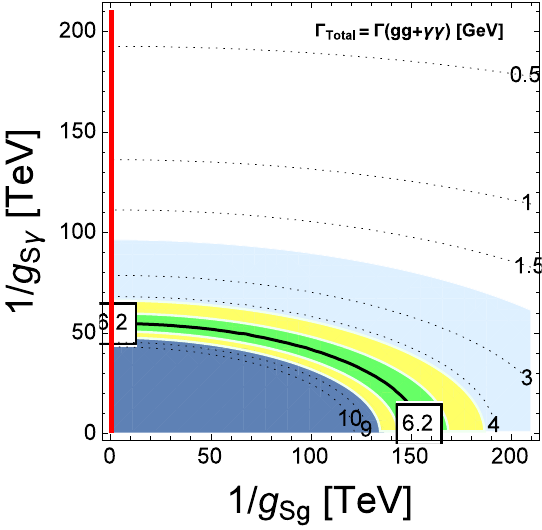}
\includegraphics[scale=0.8]{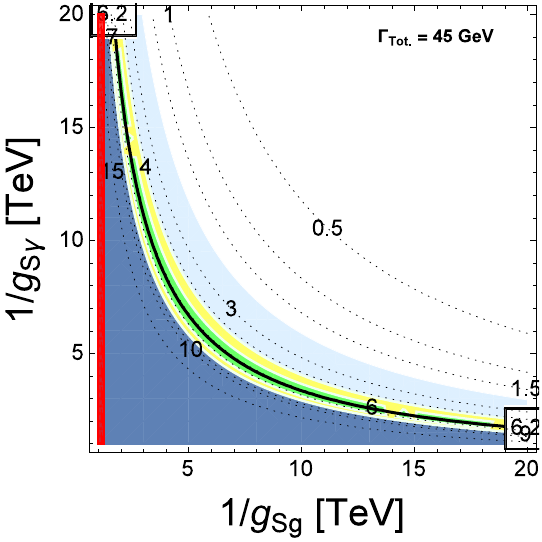}
\caption{\it Contours of production cross-section times the $\gamma \gamma$ branching ratio, in femtobarn units,
as functions of the inverses of the effective couplings in units of TeV, assuming dominance
by decays into gluon-gluon and $\gamma \gamma$ (left panel) or a total decay width of 45~GeV (right panel).
The solid black lines with green and yellow bands corresponds to the global best fit with formal 1- and 2-sigma contours. The dark and light blue shaded regions are the 68\% C.L. and 95\% C.L. bounds from CMS Run 1, respectively, and the red regions are excluded by dijet searches~\protect\cite{dijet}.  }
\label{fig:scalarEFT}
\end{figure}

The partonic $gg \to X$ production cross section has the standard leading-order expression
\begin{equation}
\hat \sigma(g g \to X) =  \frac{\pi^2}{8 m_X} \, \Gamma( X \to g g) \, \delta(\hat s -m_X^2) \, ,
\end{equation}
and this gluon-fusion production cross-section can be rescaled to the proton-proton production cross-section by numerical factors 
determined by the gluon-gluon luminosity functions at the different energies. We find that at LHC13
\bear
\sigma(g g \to X \to \gamma\gamma) &\simeq& (100 \textrm{ pb}) \times (g_{Xg}.\textrm{TeV})^2 \times \textrm{BR } (X \to \gamma \gamma) \nonumber \\
&\simeq&   (13 \textrm{ pb}) \times (g_{X\gamma}.\textrm{TeV})^2 \ ,
\label{gvsxsec}
\eear
where we have assumed that the $\gamma \gamma$ branching ratio
$BR\simeq g_{X\gamma}^2/(8 g_{Xg}^2)$, as the ratio among the couplings tends to be hierarchical:
$g_{Xg} / g_{X\gamma} \propto \alpha_s/\alpha \gg1 $.
We plot in Fig.~\ref{fig:scalarEFT} contours of the production cross-section times branching ratio in units of femtobarns,
as functions of the inverses of the effective couplings in units of TeV, for the two different decay width hypotheses. 
The solid black line denotes our best-fit cross-section of 6.2 fb, which is very compatible with the observed excess, 
while the light green (yellow) shaded region indicates 1 (2) sigma cross-sections ranging from 4.2 (5.2) to 7.2 (8.2) fb.
The $2-8$ fb bounds from Run 1 correspond to the light blue and dark blue shaded regions, and we see that the 
potential signal in Run 2 requires a cross-section that lies within this uncertainty. Given the limited statistics, 
the Run 1 and Run 2 data are quite compatible. We also show shaded in red the excluded region from dijet searches for decays into gluons~\cite{dijet}, which only places weak limits on $1/g_{Sg} \lesssim 1.2$ TeV~\footnote{The dijet limit is
 obtained from the octet scalar limit in~\cite{dijet} rescaled to 13 TeV with an acceptance of $\sim 60$\%.}.

 In the following Section we consider various models with loops of vector-like fermions to generate the 
 EFT coefficients $g_{Xg}$ and $g_{X\gamma}$, which we parametrize as a sum over vector-like fermions 
 $\psi$ with mass $m_F$ and charge $Q_F$:
\begin{eqnarray} 
{\cal L}_F = & & i \lambda_S \,  S\, \bar \psi \psi + i \lambda_P \, P \, \bar \psi \gamma^5 \psi + Q_F \, e \, \bar \psi \gamma^\mu \psi A_\mu + C_\psi g_s \, \bar \psi \gamma^\mu  \psi G_\mu \ ,
\end{eqnarray} 
where $C_\psi =1,$ 0 for coloured (un-coloured) fermions. 
The contributions to the couplings of $X=S$, $P$ to gluons can be computed by evaluating a simple
fermion loop. The resulting coupling is proportional to the trace and axial anomaly for $S$~\cite{anomaly,djouadi,radion}
and $P$~\cite{alp-decay}, respectively. 
In the next Section we present a set of models involving vector-like fermions and evaluate their effect on the 
diphoton signal as well as decays into other vector states, $WW$, $ZZ$ and $Z\gamma$, using their matching to the EFT. 

For example, in the scalar case, the contribution of a single heavy coloured fermion $F$ with charge $Q_F$
to the EFT coefficient is as follows:
\begin{eqnarray} 
\Gamma(S\to g g )& \simeq & \frac{m_S^3}{1296 \pi^3} \, \frac{N_c^2 \lambda_S^2 \alpha_s^2}{m_F^2} \, \Rightarrow  g_{Sg}=\frac{N_c \lambda_S \alpha_s}{9 \pi \sqrt{2} m_F}, \nonumber \\
\Gamma(S\to \gamma\gamma )& \simeq & \frac{m_S^3}{288 \pi^3} \, \frac{\lambda_S^2 N_c^2 Q_F^4 \alpha^2}{m_F^2} \Rightarrow g_{S\gamma}= \frac{\sqrt{2} N_c \lambda_S Q_F^2 \alpha}{3 \pi m_F} \ . 
\end{eqnarray} 
By inspecting the expression above and the total cross section at LHC13 in Eq.(\ref{gvsxsec}), 
one can see that in order to get a cross section in the region of few fb with such a single coloured fermion one would require
\bear
\lambda_S Q^2 \frac{\textrm{TeV}}{m_F} \simeq {\cal O}(\text{few}) \ ,
\eear
which indicates that this minimal scenario would require large couplings and/or a sub-TeV vector-like
fermion. In more realistic vector-like fermion models, such as those described in the next Section, 
we expect more fermionic degrees of freedom to contribute to the production, which would then scale as 
 \bear
\hat \sigma \propto  (N_F \lambda_S \alpha_s)^2 \times \frac{m_X^2} {m_F^2} \ , 
 \eear
where $N_F$ is the number of coloured fermions in the model. Moreover, the branching ratio to 
diphotons could be affected by the presence of new bosonic degrees of freedom. For example, 
one could think of incorporating the reported excess in massive dibosons at 2 TeV invariant mass~\cite{2TeV} 
within this framework. This or any other massive $W'$ would contribute to the decay of $X\to \gamma\gamma$,
but not to the $Xgg$ coupling.

\section{Models with Massive Vector-Like Fermions}

\subsection{Specifications of the Models}

Having established the general viability of models in which loops of vector-like fermions generate $X$ production and
its decay into $\gamma \gamma$, we now present four specific models, with the aim of studying their specific features, constraints
and signatures that could serve to distinguish them. As already mentioned, in all these models we assume that the $X$
particle is an isosinglet.

\textbf{Model 1:}
\label{Model1.SEC}

In Model 1, we couple the $X$ to an SU(2)-singlet \VL\ top-like quark. We define this top-like quark in two-component notation as
\begin{equation}
T \equiv \begin{pmatrix} (T_L)_\alpha \\ (T_R)^{\dagger \dot{\alpha}}\end{pmatrix} \, ,
\end{equation}
which is to be compared with the $t_R$ in the SM:
\begin{equation}
t_{R,SM} \equiv \begin{pmatrix}  0 \\ (t_R)^{\dagger \dot{\alpha}}\end{pmatrix} \, .
\end{equation}
The charge and representation assignments in this model are shown in Table~\ref{Model1Qs.TAB}.

\begin{table}[h]
\centering
\begin{tabular}{c c c c}
\hline
& \textbf{U(1)$_{em}$} & \textbf{SU(2)} & \textbf{SU(3)} \\
\hline
$X$ & 0 & 1 & 1 \\
$T_R$ & $-2/3$ & 1 & $\bar{3}$ \\
$T_L$ & $+2/3$ & 1 & 3 \\
\hline
\end{tabular}
\caption{Charge and representation assignments for the new fields in Model 1 in two-component notation.}
\label{Model1Qs.TAB}
\end{table}

Because of this choice of charges and representations, the SU(2)-singlet top-like quark can also couple via the SM Higgs field to all the
left-handed SM charge 2/3 quarks, and via bilinear mass terms to all the right-handed SM charge 2/3 quarks, since no symmetry forbids these couplings. 
Assuming mixing to only the third generation of SM quarks, the Lagrangian is then
\begin{align}
\mathcal{L} = &-\lambda_{STT} S \bar{T} T -\lambda_{PTT} P \bar{T} \gamma^5 T- M_T \bar{T} T - (\lambda_{HtT} \tilde{H} \bar{t}_L T + \mu_{tT}\overline{t}_{R,SM} T+ h.c.) \\ \nonumber &+ \mathcal{L}_{gauge~int.} + \mathcal{L}_{kinetic}\, ,
\end{align}
where $\tilde{H} = i \sigma_2 H^*$.

The mass matrix for mixing between the \VL\ states and the SM states can be written down in four-component notation as
\begin{align}
\mathcal{L} = \begin{pmatrix} \bar{t}_L & \bar{T} \end{pmatrix} \begin{pmatrix} \tilde{m}_t & \tilde{m}_{tT} \\ 0 & M_T \end{pmatrix} \begin{pmatrix} t_R \\ T \end{pmatrix} \, ,
\end{align}
where we have defined $\lambda_{i} v / \sqrt{2} = \tilde{m}_i$ with the appropriate Yukawa couplings in each case. 
We have used the fact that the $\mu_{tT}$ mass term can be rotated away by choosing a field
basis with an appropriate combination of $t_R$ and $T_R$, and redefining the Yukawa couplings.
This mass matrix is diagonalised by 
\begin{align}
\begin{pmatrix} c_{\theta_L} & -s_{\theta_L} \\ s_{\theta_L} & c_{\theta_L} \end{pmatrix}\begin{pmatrix} \tilde{m}_t & \tilde{m}_{tT} \\ 0 & M_T \end{pmatrix} \begin{pmatrix} c_{\theta_R} & s_{\theta_R} \\ -s_{\theta_R} & c_{\theta_R} \end{pmatrix} = \begin{pmatrix} m_t & 0 \\ 0 & M'_T \end{pmatrix} \, ,
\end{align}
where
\begin{align}
\tan(2\theta_L)= \frac{2 M_T \tilde{m}_{tT}}{M_T^2 - \tilde{m}_t^2 - \tilde{m}_{tT}^2}, ~~\tan(2\theta_R)= \frac{2 \tilde{m}_t \tilde{m}_{tT}}{ M_T^2 - \tilde{m}_t^2 +\tilde{m}_{tT}^2} \, .
\end{align}
For simplicity, we consider here the limit of small mixing.



\vspace{0.5cm}

\textbf{Model 2:}
\label{Model2.SEC}

In Model 2, we couple the $X$ to an SU(2)-doublet \VLQ\ partner, defined in two-component notation as
\begin{equation}
Q \equiv \begin{pmatrix}\begin{pmatrix} (U_L)_\alpha \\ (U_R)^{\dagger \dot{\alpha}}\end{pmatrix} \\ \begin{pmatrix} (D_L)_\alpha \\ (D_R)^{\dagger \dot{\alpha}}\end{pmatrix} \end{pmatrix} \, ,
\end{equation}
which may be be compared to a typical left-handed SM quark doublet:
\begin{equation}
q_{L,SM} \equiv \begin{pmatrix}\begin{pmatrix} (u_L)_\alpha \\ 0\end{pmatrix} \\\begin{pmatrix} (d_L)_\alpha \\ 0\end{pmatrix} \end{pmatrix}.
\end{equation}
The charge and representation assignments in this model are shown in Table~\ref{Model2Qs.TAB}.

\begin{table}[h]
\centering
\begin{tabular}{c c c c c}
\hline
& \textbf{U(1)$_{em}$} & \textbf{SU(2)} & \textbf{SU(3)} \\
\hline
$X$ & 0& 1 & 1 \\
$U_R$ & $-2/3$ & $\bar{2}$ & $\bar{3}$ \\
$U_L$ & $+2/3$ & 2 & 3 \\
$D_R$ & $+1/3$ & $\bar{2}$ & $\bar{3}$ \\
$D_L$ & $-1/3$ &2 & 3 \\
\hline
\end{tabular}
\caption{Charge and representation assignments for the new fields in Model 2 in two-component notation.}
\label{Model2Qs.TAB}
\end{table}

Because of this choice of charges, the SU(2)-doublet \VLQ\ can also couple via the SM Higgs field to the
right-handed SM quarks, and via a bilinear mass term to the left-handed SM quarks, since no symmetry forbids these couplings. The Lagrangian is then
\begin{align}
\mathcal{L} &= -\lambda_{SQQ} S \bar{Q} Q -\lambda_{PQQ} P \bar{Q} \gamma^5 Q - M_Q \bar{Q} Q \\&\nonumber- (\lambda_{Qt} \tilde{H} \bar{U} t_R + \lambda_{Qb} H \bar{D} b_R + \mu_{Qq}\bar{U} t_L+\mu_{Qq}\bar{D} b_L + h.c.) \\&\nonumber+ \mathcal{L}_{gauge~int.} + \mathcal{L}_{kinetic}\, .
\end{align}
As in the singlet \VLQ\ case, the bilinear mass term $\mu_{Qq}$ can be rotated away by choosing a basis
with an appropriate combination of the quark fields and redefinitions of Yukawa couplings.

The mass matrix can then be written as 
\begin{align}
\mathcal{L} = \begin{pmatrix} \bar{t}_L & \bar{U} \end{pmatrix} \begin{pmatrix} \tilde{m}_t & 0 \\ \tilde{m}_{Qt} & M_Q \end{pmatrix} \begin{pmatrix} t_R \\ U \end{pmatrix} + \begin{pmatrix} \bar{b}_L & \bar{D} \end{pmatrix} \begin{pmatrix} \tilde{m}_b & 0 \\ \tilde{m}_{Qb} & M_Q \end{pmatrix} \begin{pmatrix} b_R \\ D \end{pmatrix} \ .
\end{align}
The mass matrices can be diagonalised in the following way:
\begin{align}
\begin{pmatrix} c_{\theta^u_L} & -s_{\theta^u_L} \\ s_{\theta^u_L} & c_{\theta^u_L} \end{pmatrix}\begin{pmatrix} \tilde{m}_t & 0 \\ \tilde{m}_{Qt} & M_Q \end{pmatrix} \begin{pmatrix} c_{\theta^u_R} & s_{\theta^u_R} \\ -s_{\theta^u_R} & c_{\theta^u_R} \end{pmatrix} = \begin{pmatrix} m_t & 0 \\ 0 & M'_U \end{pmatrix}\ ,
\end{align}
and similarly for the down-type quarks:
\begin{align}
\begin{pmatrix} c_{\theta^d_L} & -s_{\theta^d_L} \\ s_{\theta^d_L} & c_{\theta^d_L} \end{pmatrix}\begin{pmatrix} \tilde{m}_b & 0 \\ \tilde{m}_{Qb} & M_Q \end{pmatrix} \begin{pmatrix} c_{\theta^d_R} & s_{\theta^d_R} \\ -s_{\theta^d_R} & c_{\theta^d_R} \end{pmatrix} = \begin{pmatrix} m_b & 0 \\ 0 & M'_D \end{pmatrix} \, ,
\end{align}
where
\begin{align}
\tan(2\theta^{u(d)}_R)= \frac{2 M_Q \tilde{m}_{Qt(b)}}{M_Q^2 - \tilde{m}_{t(b)}^2 - \tilde{m}_{Qt(b)}^2}, ~~\tan(2\theta^{u(d)}_L)= \frac{2 \tilde{m}_{t(b)} \tilde{m}_{Qt(b)}}{ M_Q^2 - \tilde{m}_{t(b)}^2 +\tilde{m}_{Qt(b)}^2} \, .
\end{align}
As before, for simplicity, we consider here the limit of small mixing.

\vspace{0.5cm}
\textbf{Model 3:}
\label{Model3.SEC}

In Model 3 we take a \VL\ copy of one generation of SM quarks, i.e., one SU(2) doublet and two singlets, with SM-like charge assignments. 
We then have a combination of the singlet \VL\ top quark defined in Section \ref{Model1.SEC}, the doublet defined in Section \ref{Model2.SEC},
and a down-type singlet \VL\ bottom quark, which can be written in two-component notation as:
\begin{equation}
B \equiv \begin{pmatrix} (B_L)_\alpha \\ (B_R)^{\dagger \dot{\alpha}}\end{pmatrix} \, ,
\end{equation}
to be compared with the right-handed SM bottom quark
\begin{equation}
b_{R,SM} \equiv \begin{pmatrix}  0 \\ (b_R)^{\dagger \dot{\alpha}}\end{pmatrix} \, .
\end{equation}
The charge and representation assignments in this model are shown in Table~\ref{Model3Qs.TAB}.

\begin{table}[h]
\centering
\begin{tabular}{c c c c c}
\hline
& \textbf{U(1)$_{em}$} & \textbf{SU(2)} & \textbf{SU(3)} \\
\hline
$X$ & 0& 1 & 1 \\
$U_R$ & $-2/3$ & $\bar{2}$ & $\bar{3}$ \\
$U_L$ & $+2/3$ & 2 & 3 \\
$D_R$ & $+1/3$ & $\bar{2}$ & $\bar{3}$ \\
$D_L$ & $-1/3$ &2 & 3 \\
\hline
$T_R$ & $-2/3$ & 1 & $\bar{3}$ \\
$T_L$ & $+2/3$ & 1 & 3 \\
\hline
$B_R$ & $+1/3$ & 1 & $\bar{3}$ \\
$B_L$ & $-1/3$ & 1 & 3 \\
\hline
\end{tabular}
\caption{Charge and representation assignments for the new fields in Model 3 in two-component notation.}
\label{Model3Qs.TAB}
\end{table}

Although there is no symmetry forbidding bilinear mass terms coupling the \VL\ SU(2) doublet to the SM doublet, 
and likewise coupling the \VL\ SU(2) singlet to the SM singlet, these mass terms can be rotated away as we saw in the previous models. 
Therefore for notational ease, we drop those terms in the Lagrangian for Model 3. We do, however, now have couplings that mix the \VL\ doublet with the \VL\ singlet via the SM Higgs boson. 
The Lagrangian for this model is then:
\begin{align}
\mathcal{L} &= -\lambda_{SQQ} S \bar{Q} Q -\lambda_{PQQ} P \bar{Q} \gamma^5 Q -\lambda_{STT} S \bar{T} T -\lambda_{PTT} P \bar{T} \gamma^5 T-\lambda_{SBB} S \bar{B} B -\lambda_{PBB} P \bar{B} \gamma^5 B\\ \nonumber &- M_Q \bar{Q} Q - M_T \bar{T} T - M_B \bar{B} B -(\lambda_{QT}\tilde{H}\bar{U} T +\lambda_{QB}H\bar{D} B + h.c.) \\&\nonumber- (\lambda_{Qt} \tilde{H} \bar{U} t_R + \lambda_{Qb} H \bar{D} b_R + \lambda_{tT} \tilde{H} \bar{t}_L T + \lambda_{bB} H \bar{b}_L B + h.c.) \\&\nonumber+ \mathcal{L}_{gauge~int.} + \mathcal{L}_{kinetic} \, .
\end{align}
The mass matrix can then be written as 
\begin{align}
\mathcal{L} = \begin{pmatrix} \bar{t}_L & \bar{T} & \bar{U} \end{pmatrix} \begin{pmatrix} \tilde{m}_t & \tilde{m}_{tT} & 0\\ 0 & M_T & \tilde{m}_{QT} \\ \tilde{m}_{Qt} & \tilde{m}_{QT} & M_Q \end{pmatrix} \begin{pmatrix} t_R \\ T \\ U \end{pmatrix} + \begin{pmatrix} \bar{b}_L & \bar{B} & \bar{D} \end{pmatrix} \begin{pmatrix} \tilde{m}_b & \tilde{m}_{bB}& 0 \\ 0 & M_B & \tilde{m}_{QB} \\ \tilde{m}_{Qb} & \tilde{m}_{QB} & M_Q \end{pmatrix} \begin{pmatrix} b_R \\ B \\ D \end{pmatrix} \, ,
\end{align}
which can be diagonalised to find the mass eigenstates. In the limit where $\tilde{m}_{bB,tT,Qb,Qt} \ll M_{B,T,Q}$, 
the \VL\ quarks can still decay into the SM quarks, and precision constraints are no longer relevant. 
Since we require only that the couplings be large enough for the decay to occur promptly, 
we assume that our model lives in this regime. 
Then we are most interested in the mass eigenstates of the \VL\ quarks themselves,
taking into account the couplings $\tilde{m}_{QT},~\tilde{m}_{QB}$. The mass matrices can then be written as
\begin{align}
\mathcal{L} = \begin{pmatrix} \bar{T} & \bar{U} \end{pmatrix} \begin{pmatrix}  M_T & \tilde{m}_{QT} \\  \tilde{m}_{QT} & M_Q \end{pmatrix} \begin{pmatrix} T \\ U \end{pmatrix} + \begin{pmatrix} \bar{B} & \bar{D} \end{pmatrix} \begin{pmatrix}  M_B & \tilde{m}_{QB} \\  \tilde{m}_{QB} & M_Q \end{pmatrix} \begin{pmatrix} B \\ D \end{pmatrix} \, ,
\end{align}
and the mass eigenstates are then found by rotating
\begin{align}
\begin{pmatrix} c_{\theta_U} & -s_{\theta_U} \\ s_{\theta_U} & c_{\theta_U} \end{pmatrix} \begin{pmatrix}  M_T & \tilde{m}_{QT} \\  \tilde{m}_{QT} & M_Q \end{pmatrix} \begin{pmatrix} c_{\theta_U} & s_{\theta_U} \\ -s_{\theta_U} & c_{\theta_U}\end{pmatrix} 
\end{align}
and analogously for the down-type quarks, with angle $\theta_D$. The solutions for the angles are
\begin{align}
\tan(2\theta_U)= \frac{2\tilde{m}_{QT}}{M_Q - M_T},~\tan(2\theta_D)= \frac{2\tilde{m}_{QB}}{M_Q - M_B} \, ,
\end{align} 
and the mass eigenvalues are given by
\begin{align}
M_{U_1}=M_Q c_{\theta_U}^2 + M_T s_{\theta_U}^2 + 2 \tilde{m}_{QT}c_{\theta_U}s_{\theta_U},&~~~M_{U_2}=M_Q s_{\theta_U}^2 + M_T c_{\theta_U}^2 - 2 \tilde{m}_{QT}c_{\theta_U}s_{\theta_U} \, , \\
\nonumber \\ 
M_{D_1}=M_Q c_{\theta_D}^2 + M_B s_{\theta_D}^2 + 2 \tilde{m}_{QB}c_{\theta_D}s_{\theta_D},&~~~M_{D_2}=M_Q s_{\theta_D}^2 + M_B c_{\theta_D}^2 - 2 \tilde{m}_{QB}c_{\theta_D}s_{\theta_D} \, .
\end{align}

\vspace{0.5cm}
\textbf{Model 4:}
\label{Model4.SEC}

In this model we consider adding \VL\ copies of a full generation of SM fermions. The particle content is therefore the same as in
Model 3, with the addition of a doublet of \VL\ leptons and a singlet \VL\ electron partner. This model can be thought of as 
adding \VL\
pairs of $\mathbf{5}+\bar{\mathbf{5}}$ and $\mathbf{10}+\overline{\mathbf{10}}$ in the language of SU(5) grand unification. An extension, 
which we will also consider below, is to add a neutral \VL\ partner, which is a pair of singlets under SU(5). This can be thought of as adding a 
$\mathbf{16}+\overline{\mathbf{16}}$ in the language of SO(10). One motivation for adding the neutral \VL\ state is that it could provide a natural dark matter (DM) candidate if it is stable. We note that renormalization effects would typically give positive corrections to the masses of
the $\mathbf{5}+\bar{\mathbf{5}}$ and $\mathbf{10}+\overline{\mathbf{10}}$ states in these $\mathbf{16}+\overline{\mathbf{16}}$ multiplets~\footnote{On the other
hand, in a SUSY version of this scenario, the lightest supersymmetric particle would also be a natural dark matter candidate.}. Since the neutral singlet plays no role in the production of $X$ or its decay, the $\mathbf{5}+\bar{\mathbf{5}} +\mathbf{10}+\overline{\mathbf{10}}$ model is recovered by setting the $N$ couplings and mass to zero. In this case the neutral component of the doublet, $L^1$ could provide a DM candidate if it is stable.

Rather than reproduce the Lagrangian from Model 3, we write here only the terms for the lepton content of Model 4. We define the \VL\ doublet as
\begin{equation}
L \equiv \begin{pmatrix}\begin{pmatrix} (L^1_L)_\alpha \\ (L^1_R)^{\dagger \dot{\alpha}}\end{pmatrix} \\ \begin{pmatrix} (L^2_L)_\alpha \\ (L^2_R)^{\dagger \dot{\alpha}}\end{pmatrix} \end{pmatrix}
\end{equation}
and the \VL\ singlets as
\begin{equation}
E \equiv \begin{pmatrix} (E_L)_\alpha \\ (E_R)^{\dagger \dot{\alpha}}\end{pmatrix},~~~~~N \equiv \begin{pmatrix} (N_L)_\alpha \\ (N_R)^{\dagger \dot{\alpha}}\end{pmatrix} \, .
\end{equation}
The charge and representation assignments in this model are shown in Table~\ref{Model4Qs.TAB},
where $e_{L,R}$ is the third-generation charged SM lepton.

\begin{table}[h]
\centering
\begin{tabular}{c c c c c}
\hline
& \textbf{U(1)$_{em}$} & \textbf{SU(2)} & \textbf{SU(3)} \\
\hline
$X$ & 0& 1 & 1 \\
$L^1_R$ & $0$ & $\bar{2}$ & $1$ \\
$L^1_L$ & $0$ & 2 & 3 \\
$L^2_R$ & $+1$ & $\bar{2}$ & $1$ \\
$L^2_L$ & $-1$ & 2 & 1 \\
\hline
$E_R$ & $+1$ & 1 & $1$ \\
$E_L$ & $-1$ & 1 & 1 \\
\hline
$N_R$ & $0$ & 1 & $\bar{1}$ \\
$N_L$ & $0$ & 1 & 1 \\
\hline
\end{tabular}
\caption{Charge and representation assignments for the new fields in Model 4 in two-component notation.}
\label{Model4Qs.TAB}
\end{table}

We mirror our approach for the quarks by only including couplings to the third generation.
We may then write down the most general Lagrangian, again taking advantage of the fact that we can rotate away the \VL-SM mixing mass bilinear 
by an appropriate redefinition of fields and Yukawa couplings:
\begin{align}
\mathcal{L}&=\mathcal{L}_{Model~3} - \lambda_{SLL} S \bar{L}L - \lambda_{SEE} S \bar{E}E  - \lambda_{PLL} P \bar{L}\gamma^5L - \lambda_{PEE} P \bar{E}\gamma^5 E  \\ \nonumber&-M_{L}\bar{L}L - M_{E}\bar{E}E - M_N\bar{N}N - (\lambda_{LE} H \bar{L}^2E + \lambda_{LN} \tilde{H} \bar{L}^1N +h.c.) \\ \nonumber& - (\lambda_{Le}H \bar{L}^2 e_R + \lambda_{\ell E}H\bar{e}_L E + \lambda_{\ell N}\tilde{H} \bar{\nu}_L N +h.c.)\\ \nonumber& + \mathcal{L}_{gauge~int.} + \mathcal{L}_{kinetic} \, .
\end{align}
We note that, by including the neutral \VL\ singlet, one could introduce an explicit Yukawa coupling to 
give mass to the SM left-handed neutrino. 
There are very stringent bounds on this Yukawa coupling, forcing it to be $\mathcal{O}(10^{-11})$ \cite{Agashe:2014kda},
so in our analysis we assume it to vanish, and we may then write the mass matrix as
\begin{align}
\mathcal{L} = \begin{pmatrix} \bar{e}_L & \bar{E} & \bar{L}^2 \end{pmatrix} \begin{pmatrix} \tilde{m}_e & \tilde{m}_{\ell E} & 0\\ 0 & M_E & \tilde{m}_{LE} \\ \tilde{m}_{Le} & \tilde{m}_{LE} & M_L \end{pmatrix} \begin{pmatrix} e_R \\ E \\ L^2 \end{pmatrix} + \begin{pmatrix}  \bar{N} & \bar{L}^1 \end{pmatrix} \begin{pmatrix} M_N & \tilde{m}_{LN} \\ \tilde{m}_{LN} & M_L \end{pmatrix} \begin{pmatrix}  N \\L^1 \end{pmatrix} \, .
\end{align}
As for the quark sector in Model 3, we can consider the limit $M_E,~M_L \gg \tilde{m}
_{\ell E},~\tilde{m}_{Le}$ without compromising the ability of the \VL\ partners to decay promptly. 
In this limit, the mass matrices reduce to mixing only among \VL\ partners:
\begin{align}
\mathcal{L} = \begin{pmatrix} \bar{E} & \bar{L}^2 \end{pmatrix} \begin{pmatrix}  M_E & \tilde{m}_{LE} \\  \tilde{m}_{LE} & M_L \end{pmatrix} \begin{pmatrix} E \\ L^2 \end{pmatrix} + \begin{pmatrix} \bar{N} & \bar{L}^1 \end{pmatrix} \begin{pmatrix}  M_N & \tilde{m}_{LN} \\  \tilde{m}_{LN} & M_L \end{pmatrix} \begin{pmatrix} N \\ L^1 \end{pmatrix} \, .
\end{align}
The mass eigenstates are then found by rotating
\begin{align}
\begin{pmatrix} c_{\theta_E} & -s_{\theta_E} \\ s_{\theta_E} & c_{\theta_E} \end{pmatrix} \begin{pmatrix}  M_E & \tilde{m}_{LE} \\  \tilde{m}_{LE} & M_L \end{pmatrix} \begin{pmatrix} c_{\theta_E} & s_{\theta_E} \\ -s_{\theta_E} & c_{\theta_E}\end{pmatrix} \, ,
\end{align}
and analogously for the neutral leptons, with angle $\theta_N$. The solutions for the angles are
\begin{align}
\tan(2\theta_E)= \frac{2\tilde{m}_{LE}}{M_L - M_E},~\tan(2\theta_N)= \frac{2\tilde{m}_{LN}}{M_L - M_N} \, ,
\end{align} 
and the mass eigenvalues are given by
\begin{align}
M_{E_1}=M_L c_{\theta_E}^2 + M_E s_{\theta_E}^2 + 2 \tilde{m}_{LE}c_{\theta_E}s_{\theta_E},&~~~M_{E_2}=M_L s_{\theta_E}^2 + M_E c_{\theta_E}^2 - 2 \tilde{m}_{LE}c_{\theta_E}s_{\theta_E} \, , \\
\nonumber \\ 
M_{N_1}=M_L c_{\theta_N}^2 + M_N s_{\theta_N}^2 + 2 \tilde{m}_{LN}c_{\theta_N}s_{\theta_N},&~~~M_{N_2}=M_L s_{\theta_N}^2 + M_N c_{\theta_N}^2 - 2 \tilde{m}_{LN}c_{\theta_N}s_{\theta_N} \, .
\end{align}

The lighter of the two neutral leptons could be a dark matter candidate if it is stable. It is precisely this observation which leads us to have written down the couplings between the neutral \VL\ lepton and the hypothetical $X=$ $S$ or $P$ fields, because while they do not contribute to the production or decay of $S$/$P$, they would be important for the calculation of the relic density. Models involving a radion, like our $S$ particle, and axion, i.e., $P$, have been studied elsewhere, see, e.g., Refs~\cite{axion1,axion2,vero-axion, vero-radion}, and in this case the main annihilation would be to gluons:
\bear
N_1 \, N_1 \to X \to g \, g \, .
\eear
This annihilation is p-wave suppressed for the case of the scalar and s-wave for the pseudoscalar candidate. The annihilation cross section for the pseudo-scalar is given by
\bear
\langle\sigma v\rangle_{g g }&= & \, \frac{4 |\lambda_a|^4 \alpha_s^2 }{\pi^3 }\,  \cdot
\frac{m^2_{N_1}}{(4m^2_{N_1}-m^2_a)^2+\Gamma^2_a m^2_a} \, . 
\eear
We note that a large cross section for annihilation into gluons could in principle be probed in direct detection experiments, although the limits degrade steeply with the dark matter particle mass, and above 300 GeV it is out of reach of the XENON1T
that is now starting~\cite{Chu:2012qy} .

\subsection{Summary of Vector-Like Models}
\label{VLSummary.SSEC}

For the reader's convenience, we present here a short summary of each model we consider. We list in Table \ref{ModelSummary.SEC} the new field contents of the various models, 
now in four-component notation.

\begin{table}[h]
\centering
\begin{tabular}{c c c c c}
\hline
\textbf{Model} & \textbf{Field content} & \textbf{U(1)$_{em}$} & \textbf{SU(2)} & \textbf{SU(3)} \\
\hline
All models & X & 0 & 1 & 1 \\
 \hline
 \textbf{1}, \textbf{3} \& \textbf{4} & $T$ & +2/3 & 1 & 3 \\
  & $\bar{T}$ & -2/3 & 1 & $\bar{3}$ \\
  \hline
 \textbf{2}, \textbf{3} \& \textbf{4} & $U$ & +2/3 & 2 & 3 \\
  & $\bar{U}$ & -2/3 & $\bar{2}$ & $\bar{3}$ \\
 \textbf{2}, \textbf{3} \& \textbf{4} & $D$ & -1/3 & 2 & 3 \\
  & $\bar{D}$ & +1/3 & $\bar{2}$ & $\bar{3}$ \\
  \hline
 \textbf{3} \& \textbf{4} & $B$ & -1/3 & 1 & 3 \\
  & $\bar{B}$ & +1/3 & 1 & $\bar{3}$ \\
  \hline
\textbf{4} & $L^1$ & 0 & 2 & 1 \\
& $\bar{L}^1$ & 0 & $\bar{2}$ & $1$ \\
\textbf{4} & $L^2$ & -1 & 2 & 1 \\
  & $\bar{L}^2$ & +1 & $\bar{2}$ & $1$ \\
\textbf{4} & $E$ & -1 & 1 & 1 \\
  & $\bar{E}$ & +1 & 1 & $1$ \\
  \textbf{4} & $N$ & 0 & 1 & 1 \\
  & $\bar{N}$ & 0 & 1 & $1$ \\
  \hline
\end{tabular}
\caption{The new field contents of all the models under consideration, in four-component notation. }
\label{ModelSummary.SEC}
\end{table}

If we assume, for simplicity, a degenerate spectrum for each model, and universal couplings, we can easily quantify the predicted branching ratios for each decay mode of the particle $X$ as a function  of the number of fermions and their charges under $SU(2)_L\times U(1)_Y$. The couplings are as follows
\bear
g_{X\gamma} &=& c_1 \alpha_Y c_W^2 + c_2 \alpha_2 s_W^2 \, , \nonumber \\
g_{X Z \gamma } &=& ( c_1 \alpha_Y- c_2 \alpha_2) s_{2W} \, , \nonumber \\
g_{X ZZ } &=& c_1 \alpha_Y s_W^2 + c_2 \alpha_2 c_W^2 \, , \nonumber \\
g_{XWW} &=& 2 c_2 \alpha_2 \ ,
\eear
where $s_W=\sin_{\theta_W}$, $s_{2W}=\sin{2\theta_{W}}$, with $\theta_W$ the weak mixing
angle, and $\alpha_{Y,2} = g_{Y,2}^2/4 \pi$. The coefficients $c_{1,2}$ are given by
\bear
c_{1} &=& \sum_F \lambda  Tr[Y^2] f_{1/2}(\tau_F)\ , \nonumber \\
c_{2} &=& \sum_F \lambda  Tr[D(r)^2] f_{1/2}(\tau_F)\ ,
\label{definef}
\eear
where $f_{1/2}(\tau_F) $ is a triangle loop function, and $Y$ and $D(r)$ are the hypercharge and Dynkin index of 
the representation $r$ of the fermion $F$, respectively. The triangle loop function is defined as 
\begin{align}
\nonumber f_{1/2}(\tau_F) =2\left(\tau_F+\left( \tau_F-1\right)f(\tau_F)\right)\tau_F^{-2}\, , \\
 f\left(\tau_F\right) = \arcsin^2\sqrt{\tau_F}, ~~~ \tau_F \leq 1 \,
\label{fdefinition}
\end{align} 
where $\tau_F = m_X^2/4m_F^2$. In the limit we consider where $\tau_F \ll 1$, $f_{1/2}(\tau_F) \to 4/3$. The contribution to the gluon coupling can be obtained in a similar way as the other couplings. We use these expressions to obtain the ratios of partial widths to vector bosons in the various models listed in Table~\ref{tab:BR}.

\begin{table}
\centering
\begin{tabular}{c c c c c c c}
\textbf{Model} & $ Tr[Y^2]$& $ Tr[D(r)^2] $ & $\frac{\text{BR}(X\to gg)}{\text{BR}(X\to \gamma \gamma)}$ & $\frac{\text{BR}(X\to Z \gamma)}{\text{BR}(X\to \gamma \gamma)}$ & $\frac{\text{BR}(X\to Z Z)}{\text{BR}(X\to \gamma \gamma)}$ & $\frac{\text{BR}(X\to W^\pm W^\mp)}{\text{BR}(X\to \gamma \gamma)}$ \\
\hline
\textbf{1} & $8/3$ & 0 & 180 & 1.2 & 0.090 & 0\\
\textbf{2} & $1/3$ & 3  & 460 & 10 & 9.1 & 61\\
\textbf{3} & $11/3$ & 3 & 460 & 1.1 & 2.8 & 15\\
\textbf{4} & $20/3$ & 4 & 180 & 0.46 & 2.1 & 11\\
\hline
\textbf{Current limit} & &  & $\sim 2 \times 10^4$ &7 & 13 & 30
\end{tabular}
\caption{Group indices and ratios of branching ratios for the various models under consideration,
where we have used $\alpha_s (m_X) \simeq 0.092$. The upper limit on $\frac{\text{BR}(X\to gg)}{\text{BR}(X\to \gamma \gamma)}$ is obtained from the left panel of Fig.~\ref{fig:scalarEFT}, and
explanations how we derive the other current limits are provided in Section \ref{Searches.SEC}.}
\label{tab:BR}
\end{table}

\begin{figure}
\centering
\includegraphics[scale=0.75]{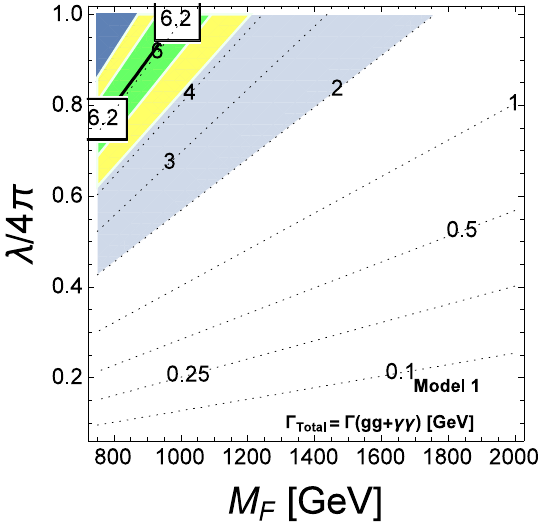}
\includegraphics[scale=0.75]{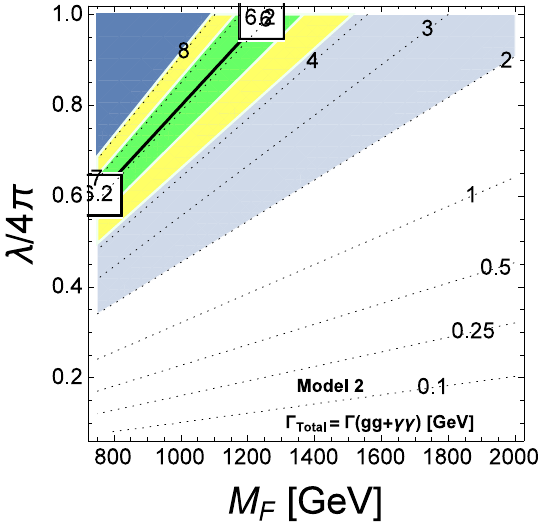} \\
\includegraphics[scale=0.75]{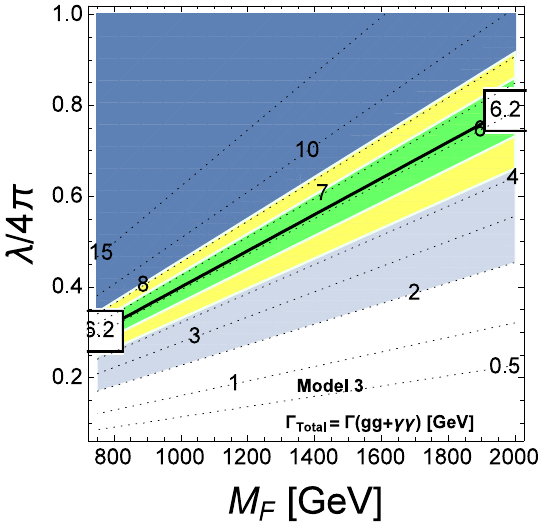}
\includegraphics[scale=0.75]{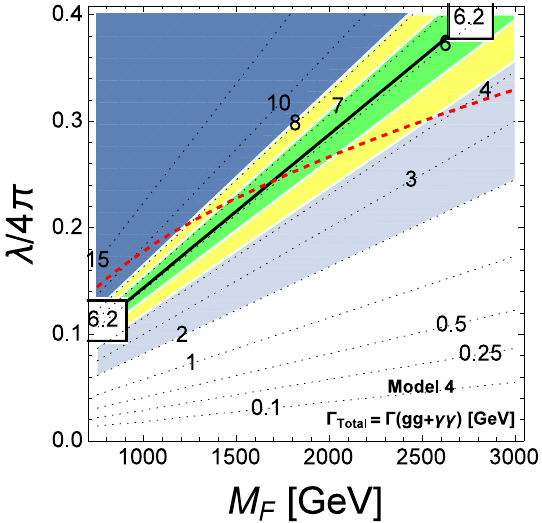}
\caption{\it Contours of production cross-section times the $\gamma \gamma$ branching ratio in femtobarn
units for the four models we consider, assuming a narrow width with decays into gluons dominating.
The shaded light and blue regions correspond to the weaker and stronger 95\% CL exclusion limits from CMS Run 1,
while the green and yellow bands represent our indicative 1- and 2-sigma ranges around the best fit 
cross-section for the tentative signal, represented by black lines. The dashed red line in the lower right panel
corresponds to the observed relic abundance~\protect~\cite{planck}.}
\label{fig:scalarmodels}
\end{figure}

\begin{figure}
\centering
\includegraphics[scale=0.75]{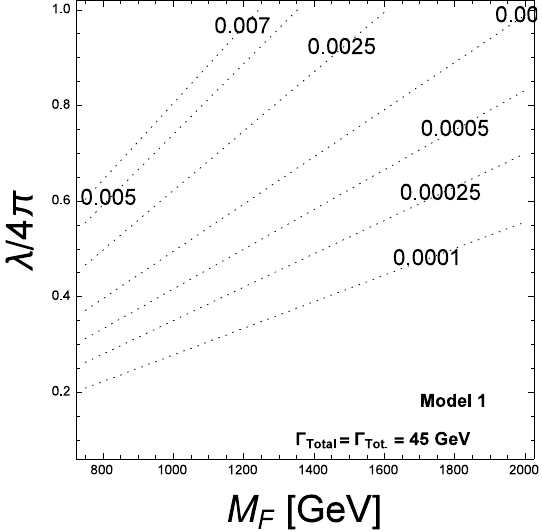}
\includegraphics[scale=0.75]{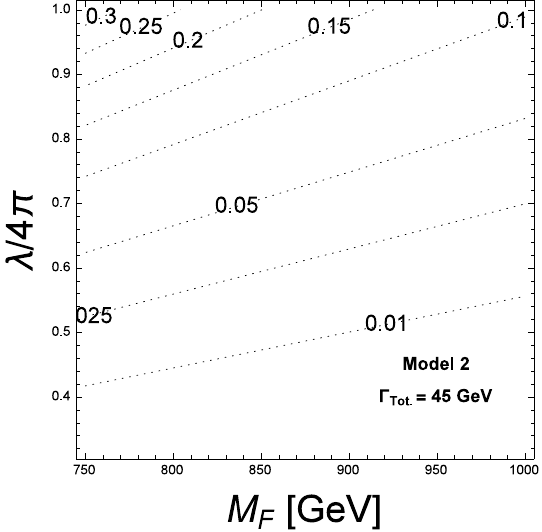} \\
\includegraphics[scale=0.75]{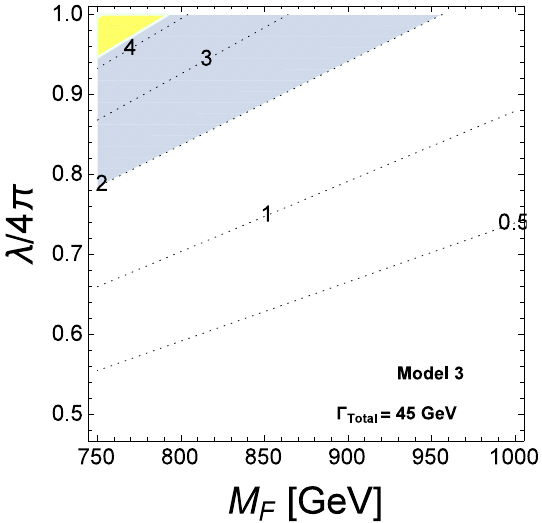}
\includegraphics[scale=0.75]{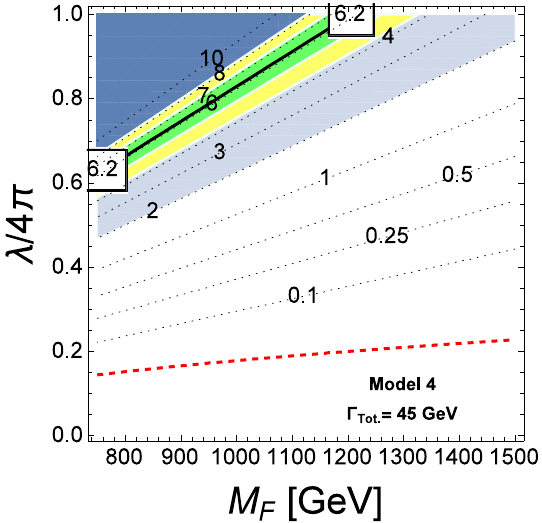}
\caption{\it Contours of production cross-section times the $\gamma \gamma$ branching ratio in femtobarn
units for the four models we consider, assuming a 45~GeV total width.
The shaded light and blue regions correspond to the weaker and stronger 95\% CL exclusion limits from CMS Run 1,
while the green and yellow bands represent our indicative 1- and 2-sigma ranges around the best fit 
cross-section for the tentative signal, represented by black lines. The dashed red line in the lower right panel
corresponds to the observed relic abundance~\protect~\cite{planck}.}
\label{fig:scalarmodelslargewidth}
\end{figure}

In Fig.~\ref{fig:scalarmodels} and Fig.~\ref{fig:scalarmodelslargewidth} we display the contours of production cross-section times $\gamma\gamma$ branching ratio, with the 1- and 2-sigma bands in light green and yellow denoting the favoured region by a global fit to the ATLAS and CMS data and the dark (light) blue regions the weakest (strongest) exclusions at 95\% CL by Run 1 of CMS. Fig.~\ref{fig:scalarmodels} assumes a photon and gluon dominated branching ratio with a narrow width, and we see that models 1 and 2 must be in a strongly-coupled and/or relatively low mass regime to obtain a large enough signal cross-section. This is alleviated somewhat in model 3 with the larger number of fermion contributions, and model 4 is a fully perturbative weakly-coupled model. 

We note, in particular, that Model 4 contains a dark matter candidate, and we show the relic density constraint~\cite{planck} by a red dashed line
in the lower right panel of Fig.~\ref{fig:scalarmodels}. For a large range of dark matter particle masses, this contour lies within the
bands favoured at the 1- and 2-$\sigma$ level.

On the other hand, Fig.~\ref{fig:scalarmodelslargewidth} assumes a large width corresponding to 6\% of the 750 GeV resonance mass~\footnote{We do not address the model-dependent issue what additional modes might dominate $X$ decays in this case.
These might be induced by small couplings to some Standard Model particles such as $\bar{t} t$, which would be allowed by experimental constraints as discussed in~\cite{DEGQ}, or there might be
invisible decays.}, which excludes all of model 1 and 2 for $\lambda < 4\pi$ and practically all of Model 3. Only model 4 survives in a corner of the parameter space with strong coupling. There is therefore a tension between increasing the decay width and perturbativity for the models we consider here. Moreover, the relic density constraint~\cite{planck} indicated by the red dashed line does not traverse the 1- and 2-$\sigma$ bands.

\subsection{Present and Future Constraints on Vector-Like Partners}

The charged \VL\ fermions are not stable, and decay via small Yukawa couplings to the Standard Model fermions via the SM Higgs boson. 
As such, a vector-like partner can have either a prompt or a displaced decay. If the decay is to be prompt, which we define as $c\tau \lesssim 100 \mu$m, 
then we can place a limit on $\lambda_{SM-VL}^2\cdot M_{VL} \gtrsim 1.6 \times 10^{-10}$ GeV~\cite{Ellis:2014dza}. 
If we assume couplings only to the third generation of SM fermions, then there are no applicable constraints due to induced tree-level FCNC decays such as $\tau \to \mu \gamma$ or $t \to Z/Hc$. The constraints in the case of mixing with the third generation arise from the oblique parameters $S$ and $T$ ($\epsilon_{1,2}$), the $Z_{b\bar{b}}$ coupling and 
the modification of $|V_{tb}|$. In the limit where mixing with the SM is small, however, these constraints no longer apply\cite{Cacciapaglia:2010vn, Cacciapaglia:2011fx}. Since our models do not need large couplings to the SM,
but just require that the decay occurs, the constraints on mixing with the SM particles are not strong in our models. 

\begin{figure}
\centering
\includegraphics[scale=0.6]{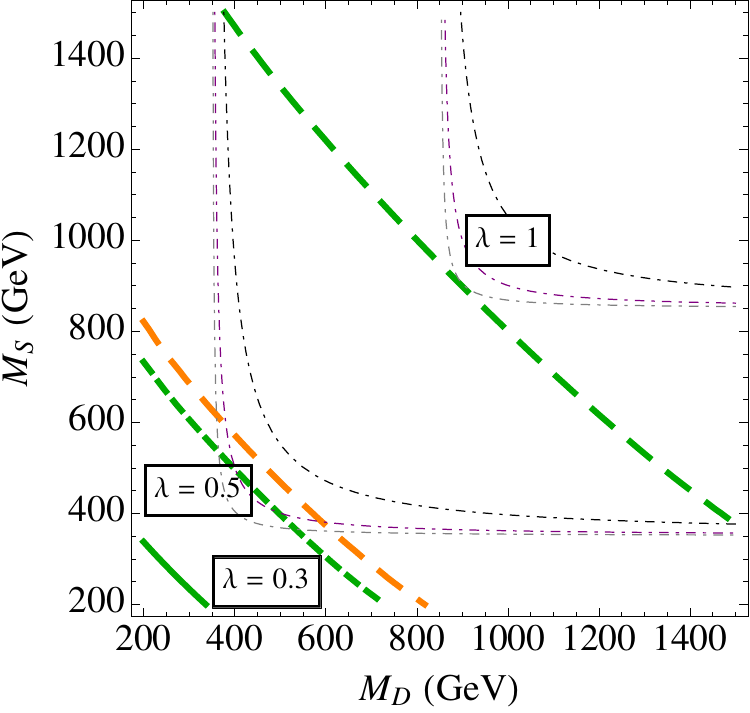}
\caption{Constraints in Model 4 on a common doublet mass $M_{D}$ and a common singlet mass $M_{S}$ from the electroweak oblique correction parameters $S$ and $T$
for various values of a common coupling $\lambda_{XY}$. We show contours for the 95 \% C.L. (green) and, in the case of $\lambda_{XY}=\lambda_{XY}=1$, 
the contour for $3\sigma$ (orange). The SM values for $S$ and $T$ lie between the 68 \% and 95 \% C.L.s. The dot-dashed contours are for constant mass of the lightest \VL\ state $M = 350,~850$ GeV for $\lambda_{XY}=0.3,0.5,1$ in grey, purple and black respectively. The choice of contours is motivated by limits on \VL\ leptons \cite{Dermisek:2014qca} and quarks \cite{Aad:2015kqa,Khachatryan:2015oba} respectively.}
\label{Model4TVtbMassConstraint.FIG}
~~ \\
\hspace{0.5cm}
\includegraphics[scale=0.6]{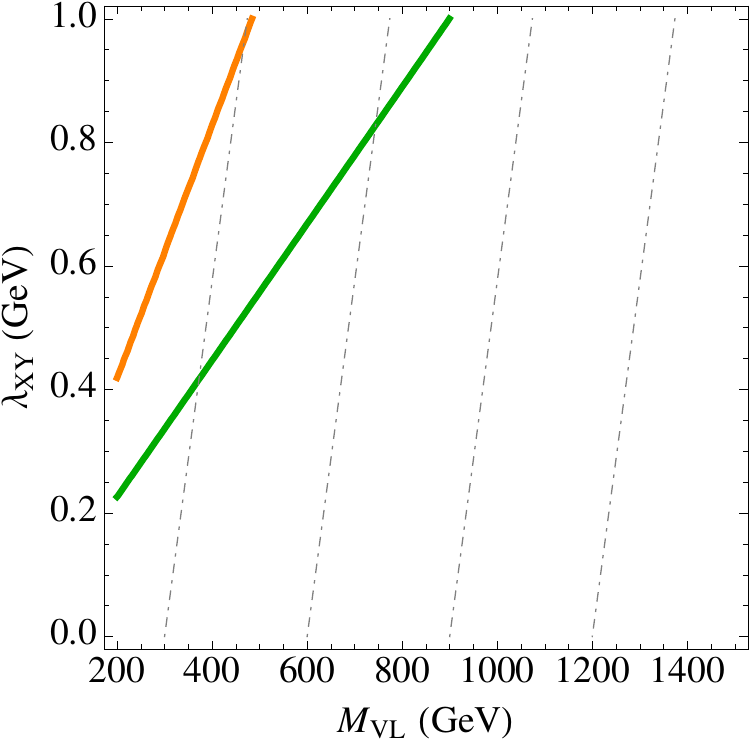}
\caption{Constraints in Model 4 on a common $\lambda_{XY}$ and $M_{VL}$ from the electroweak oblique correction parameters $S$ and $T$. We show contours for the 95 \% C.L. (green) and $3\sigma$ (orange). 
The SM values for $S$ and $T$ lie between the 68 \% and 95 \% C.L.s. The dot-dashed contours are for constant mass of the lightest \VL\ state, corresponding to $M = 300,~600,~900,~1200$ GeV.}
\label{Model4TVtbLambdaConstraint.FIG}
\end{figure}

In Models 3 and 4 however, there are relevant constraints from the electroweak oblique parameters $S$ and $T$ ($\epsilon_{1,2}$), due to mixing between the \VL\ states themselves via the SM Higgs, which we calculate using the results of \cite{Ellis:2014dza}. We show in Figs. \ref{Model4TVtbMassConstraint.FIG} and \ref{Model4TVtbLambdaConstraint.FIG} our results for Model 4. (The results for Model 3 are quite similar.) It is important to note that the central values from the {\tt GFitter} collaboration for $S$ and $T$ (after fixing $U=0$) exclude the SM at more than the 68 \% C.L.~\cite{Baak:2014ora}.
Therefore, even in the large-mass decoupling limit for the \VL\ states, the contours of regions allowed by $S$ and $T$ never drop below the 68 \% C.L. contour for either model.

Another constraint that should be taken into account is the effect of adding \VL\ fermions that mix with the SM Higgs on the Higgs couplings themselves. This has been studied in various guises (see for example \cite{Kearney:2012zi, Feng:2013mea, Dermisek:2013gta, Ellis:2014dza, Lalak:2015xea}), finding that even for relatively large mixing between the \VL\ fermions, it is possible that the Higgs couplings are not  shifted dramatically, so they can be compatible with experimental bounds.

Searches for coloured vector-like quarks have been performed at Run~1 by  ATLAS~\cite{Aad:2015kqa} 
and CMS~\cite{Khachatryan:2015oba} (\VL\ tops only) reaching about 800 GeV. 
The increase of production from 8 TeV to 13 TeV is ${\cal O}(10-20)$ for the region of 900 to 1200 GeV, 
but the backgrounds grow at a similar rate. Nevertheless, boosted techniques and more efficient 
multivariate discrimination techniques may lead to a Run 2 sensitivity to vector-like quarks around 2 TeV 
for models with coloured particles, see e.g., Ref.~\cite{Matsedonskyi:2015dns} for a recent study.
However, the current LHC limits on \VL\ quarks are already sufficient to push the fermionic form factor
$f_{1/2}(\tau_F)$ (\ref{fdefinition}) close to its asymptotic value~\footnote{The same would be true for any massive $W'$ that
might contribute to the $X\gamma\gamma$ vertex. We note that, unlike the case of the Higgs boson where
the relative signs of fermion and boson loops are opposite, the same is not necessarily the case for
their contributions to the $X\gamma\gamma$ vertex, where they may interfere constructively.}. 
The same is not necessarily the case for any \VL\ leptons, but we
assume it here, for simplicity.

\section{Other Searches for $X (750)$ at LHC Run 2}
\label{Searches.SEC}

We now recast the constraints that have been established by the ATLAS and CMS Collaborations
on diboson final states in the context of heavy SM Higgs boson searches.
We concentrate on the experimental analyses that provide the most constraining results for a state of mass $\simeq 750$~GeV.
Since we are assuming the the couplings of the X resonance to the SM fermions are small, 
we focus on possible X decays to SM gauge bosons, or to the Higgs boson, or to both of them.
With regard to the exploitation of the experimental analyses of a heavy Higgs
boson $H \rightarrow ZZ$, $WW$, we note that the vertex for an electroweak singlet decaying into a pair of gauge bosons
given in (\ref{eq:Leff}) is different from that of a Standard Model-like Higgs boson. 
In the case studied here of a CP-conserving spin-0 field, $X$, decaying into a pair of on-shell
spin-1 particles with masses much smaller than $m_X$ via an $F_{\mu \nu} F^{\mu \nu}$ or
$\epsilon^{\mu \nu \rho \sigma} F_{\mu \nu} F_{\rho \sigma}$ vertex,
there is only one possible helicity amplitude~\footnote{Processes involving at least one off-shell boson, such
as the production of the $X$ boson in association with a gauge boson, would provide good opportunities
to distinguish between Lorentz structures~\cite{Masso:2012eq}.}, yielding final states split equally
between $\pm 1$ helicity states. 
Consequently, the kinematics of such an electroweak singlet $X$ decaying to pairs of gauge bosons
should be different from the case of a heavy Higgs boson, where also zero-helicity states may be produced. 
However, we have checked that the differences in acceptance are at the 10 to 15\% level for both the $ZZ$ and $WW$
final states, and are not important for our purposes.

Limits can be borrowed from searches for a heavy SM Higgs boson in its decays to massive gauge bosons 
$X \rightarrow ZZ$.
The search for $H \rightarrow ZZ \rightarrow 4 l $, and $H \rightarrow ZZ^{*} \rightarrow 2l 2q, 2l 2\nu, 2l 2\tau$ channels,  have been performed in the framework of the SM with the full event sample recorded at the LHC run 1, namely 5.1 fb$^{-1}$ at $\sqrt{s}=7$ TeV and 19.7 fb$^{-1}$ at 
$\sqrt{s}=8$ TeV for CMS~\cite{CMSWWZZ} and 20.3 fb$^{-1}$ at 
$\sqrt{s}=8$ TeV for ATLAS~\cite{ATLASZZ}. The mass range analyzed extended to $M_{X}=1$~TeV. One should note that in a dedicated X search, this $X \rightarrow ZZ$ channel will lead to more effective constraints as heavy SM Higgs particles have total decay widths that are completely different, {\it a priori}. Whereas the SM state would have been a very wide resonance (for a mass $\simeq 750$~GeV the total decay width is $\simeq 250$~GeV), the $X$ boson might be a relatively narrow resonance as discussed previously, allowing one to select smaller bins for the $ZZ$ invariant masses that lead to a more effective suppression of the backgrounds.
CMS expressed their result in term of a ratio between the number of observed events relative to the SM expectation.
Translated into cross-sections, the observed $95\%$CL limit for a $750$~GeV SM-like resonance reads:  
$\frac{\sigma^{\text{limit}}}{\sigma_{SM}} \simeq 0.5$. This experimental limit gives then a upper value on the production cross section of the $X$ particle decaying to $Z$ bosons during LHC Run 1 that is
\begin{equation*}
\sigma(gg \rightarrow X) \times \text{BR}(X \rightarrow ZZ) \lesssim 23~{\rm fb} \, ,
\end{equation*}
which can be re-written in the form $\sigma(gg \rightarrow X \rightarrow \gamma\gamma) \times \frac{\text{BR}(X \rightarrow ZZ)}{\text{BR}(X \rightarrow \gamma\gamma)} \lesssim 23$~fb. We have seen previously that LHC Run 1 put a upper limit of order 1 fb for the $X$ production cross section times its branching ratio to photons. Therefore, we end up with a first crude estimate that 
$\frac{\text{BR}(X \rightarrow ZZ)}{\text{BR}(X \rightarrow \gamma\gamma)} \lesssim 23$.
ATLAS results give the $95\%$CL upper limit $\sigma(gg \rightarrow X) \times \text{BR}(X \rightarrow ZZ) \lesssim 13$~fb, which translates into a slightly better limit $\frac{\text{BR}(X \rightarrow ZZ)}{\text{BR}(X \rightarrow \gamma\gamma)} \lesssim 13$.

Similarly to what has been done before, one can borrow the constraint from searches for a heavy SM Higgs boson via
its decays to $W$ bosons\cite{ATLASWW,CMSWWZZ} in order to put a constraint on the decay $X \rightarrow W^{\pm}W^{\mp}$.
Searches for the $H \rightarrow W^{\pm}W^{\mp} \rightarrow 2l 2\nu, l \nu 2q$ channels have been performed in the framework of the SM with the full event sample recorded at the LHC run 1, namely 20.3 fb$^{-1}$ at $\sqrt{s}=8$ TeV for $H \rightarrow WW^{*}$ \cite{ATLASWW} in the case of ATLAS, where the high mass range was analyzed.

As we noted in the $X \rightarrow ZZ$ case, one should perform an optimized $X$ search,
since a heavy SM Higgs state would be very wide, whereas the $X(750)$ boson could be a
much narrower resonance, allowing one to select
smaller bins for the $WW$ invariant masses that lead to a more effective suppression of 
the various backgrounds.

The observed ATLAS $95\%$CL limit for a $750$~GeV SM-like resonance decaying into two $W$ bosons gives
an upper value on the production cross section of the X particle decaying to W bosons during LHC Run 1 that is  
$\sigma(gg \rightarrow X) \times \text{BR}(X \rightarrow W^{\pm}W^{\mp}) \lesssim 30$~fb. This limit assumes a gluon fusion production mode and a signal with a narrow width.
Since LHC Run 1 put a upper limit of order 1 fb for the $X$ production cross section times its branching ratio to photons, 
we end up with the crude estimate that $\frac{\text{BR}(X \rightarrow W^{\pm}W^{\mp})}{\text{BR}(X \rightarrow \gamma\gamma)} \lesssim 30$.

Searches for a narrow width resonant $X \rightarrow hh$ channel have been conducted by both the ATLAS and CMS collaborations 
with the $\approx 20$ fb$^{-1}$ of data collected at $\sqrt{s}=8$ TeV. They focused on the $\gamma\gamma b\bar{b}$
signature \cite{ATLAShh,CMShh} and also on the 4 $b$--quark final state~\cite{ATLAS4b,CMS4b}. The latter is the most constraining.
ATLAS and CMS obtained similar $95\%$CL limits for a $750$~GeV SM-like resonance decaying into two 125~GeV Higgs,
namely $\sigma(pp \rightarrow X \rightarrow h h) = 41$~fb. 
This may be translated into an upper bound on the ratio between $X$ decays to two SM Higgs bosons and to photons, 
$\frac{\text{BR}(X \rightarrow hh)}{\text{BR}(X \rightarrow \gamma\gamma)} \lesssim 41$.

The $A\rightarrow Zh \rightarrow 2l b\bar{b}$ search performed by the ATLAS and CMS collaborations also constrains the ratio  
$\frac{\text{BR}(X \rightarrow Zh)}{\text{BR}(X \rightarrow \gamma\gamma)},$ if the $X$ particle is a pseudoscalar $P$.
Unfortunately, the CMS analysis that considered the final state $2l b\bar{b}$ with 
the $\approx 20$ fb$^{-1}$ collected at $\sqrt{s}=8$~TeV \cite{CMShZ} does not cover the range $M_{X} \geq 600$~GeV. 
However for the mass range of interest, the ATLAS collaboration did a seach for $A \rightarrow Zh$ 
with the SM Higgs decaying to either a pair of bottom quark or a tau lepton pair 
and the $Z$ boson decaying to an electron pair, muon pair or neutrinos 
(in this last case the Higgs boson is required to decay into a bottom quark pair). The analysis has been done 
with the $20.3$ fb$^-1$ collected at the $\sqrt{s}=8$~TeV run~\cite{ATLAShZ}. For a pseudoscalar resonance with 
$M_{X}=750$~GeV, produced through gluon fusion, an upper limit of  $\sigma(X\rightarrow Zh) = 2 \times 10^{-2}$~pb has been set at the $95\%$ C.L. on the total production rate. 
We infer that, if the X particle is a pseudoscalar particle, its decay to the SM Higgs particle and a $Z$ boson should satisfy the requirement $\frac{\text{BR}(X \rightarrow Zh)}{\text{BR}(X \rightarrow \gamma\gamma)} \lesssim 20$.

Finally, the ATLAS Collaboration has searched for new resonances decaying to final states with a $Z$ vector boson produced in association with a high transverse momentum photon, $Z\gamma$. The measurements use $20.3$~fb$^{-1}$ of recorded data at a centre-of-mass energy of $\sqrt{s} = 8$~TeV~\cite{ATLASZgamma}. They set an upper limit of the order $7$~fb on the $\sigma(pp \rightarrow X \rightarrow Z\gamma)$ cross section. This gives the limit $\frac{\text{BR}(X \rightarrow Z\gamma)}{\text{BR}(X \rightarrow \gamma\gamma)} \lesssim 7$.

Comparison of the limits discussed in the paragraphs above with the model calculations in Table~\ref{tab:BR}
indicates that Model 2 could already be ruled out on the basis of $\frac{\text{BR}(X\to Z \gamma)}{\text{BR}(X\to \gamma \gamma)}$
and $\frac{\text{BR}(X\to W^\pm W^\mp)}{\text{BR}(X\to \gamma \gamma)}$. However, in view of the inevitable
uncertainties in recasting the LHC upper limits in these cases, we would not regard this conclusion as definitive.
Certainly, none of the other models can yet be excluded.

Until now, we have assumed in this analysis a small mixing between the new \VL\ states and the SM fermionic fields,
but LHC Run 1 data allow us to derive constraints on the couplings between the $X$ particle and SM fermions
such as the tau lepton and the top quark, which we summarize now.

- Using the ATLAS and CMS Run 1 searches for a heavy SM-like Higgs scalar decaying into a pair of tau leptons~\cite{ATLAStautau,CMStautau}, 
one can derive the following upper limit on the X coupling to tau leptons:
$\frac{\text{BR}(X \rightarrow \tau\tau)}{\text{BR}(X \rightarrow \gamma\gamma)} \lesssim 20$.

- The search for resonances decaying into $t\bar t$  final states 
will be mandatory in order to probe the potential coupling to SM fermions. 
However,  a peak in the invariant mass distribution of the $t\bar t$ system, 
that one generally expects to be quite narrow in our framework, is not the only signature of
a scalar resonance in this case. Indeed, the  $gg\to X$ signal will interfere with the QCD  
$t\bar t$ background, which is mainly generated by the gluon-fusion channel, $gg \rightarrow t\bar t$,
within the energy range of the LHC~\cite{Htta}. The interference between the signal and background will depend on the CP nature of 
the $X$ particle and on its width, see for instance~\cite{Httb,Httc,Httd}. These interferences could be either destructive or constructive, leading to a rather sophisticated signature with a ``peak and dip" structure of the $t\bar t$ invariant mass distribution. 
The $t\bar{t}$ background in the SM is known to be difficult to deal with. However, if the width of the new resonance is narrow the experimental analysis should be able to select a smaller bin size for the $t\bar{t}$ invariant masses that would lead to a more effective suppression of the backgrounds.
The ATLAS collaboration has performed a search for a spin-0 scalar color singlet resonance in the $t\bar{t}$ final state via gluon fusion using 
lepton-plus-jets events \cite{ATLAStt} . This analysis used the 20.3 fb$^{-1}$ collected at a centre-of-mass energy of 8 TeV. 
Interference between the QCD process and SM $t\bar{t}$ production has not been considered in this study. 
However, as a first attempt, one could still
use this analysis to constrain the ratio between the $X$ decays into a top quark pair and its decays to photons. 
The upper limit at $95\%$CL on the total production rate is $\sigma(p p \rightarrow X \rightarrow t \bar{t}) \leq 0.6$~pb. 
We therefore deduce that $\frac{\text{BR}(X \rightarrow t\bar{t})}{\text{BR}(X \rightarrow \gamma\gamma)} \lesssim 600$.


\section{Conclusions}

Although the enhancements reported by CMS and ATLAS in their $\gamma \gamma$ spectra around 750~GeV
are very suggestive, it remains to be seen whether the reported signal will survive as the integrated luminosity of Run~2 of
the LHC increases. Until its fate is clear, however, while maintaining due caution in view of the inconclusive significance of
the signal as well as its angular and energy dependence, it is appropriate to consider 
possible interpretations, with the objective of identifying experimental signatures that could help clarify its origin.

We have focused in this paper on possible interpretations of the signal as a spin-zero $X(750)$~GeV state decaying
into $\gamma \gamma$ that is produced via gluon-gluon fusion. We assume that the $X g g$ and $X \gamma \gamma$ 
vertices are generated by loops of heavy fermions and charged bosons, as is the case of the SM Higgs boson. However,
the fermions coupled to the $X(750)$~GeV state must have masses $\gtrsim m_X$: the heaviest known fermions
and charged bosons $t$ and $W^\pm$ could not make significant contributions. Accordingly, we have postulated
the existence of \VL\ fermions. 

We have shown that a single heavy \VLQ\ could explain the data only if its coupling to the $X(750)$ state were
close to the limit of validity of perturbation theory (which might be understandable in some strongly-coupled
composite model) and if the total decay width is not too large. However, a smaller coupling would be sufficient if the $X g g$ and $X \gamma \gamma$ loops
featured more \VL\ fermions, or if there was a contribution to the $X \gamma \gamma$ vertex from heavy bosons.

We have considered various \VL\ fermion models, ranging from a single \VL\ quark to a complete 
$\mathbf{16}+\overline{\mathbf{16}}$ pair of multiplets.
All these models would predict $X \to Z Z$ and $Z \gamma$ decays at characteristic rates relative to $X \to \gamma \gamma$,
and some models also predict $X \to W^- W^+$ decays via loop diagrams. As we have shown, 
the predicted signals from these additional
$X$ decays are compatible with the available upper limits on massive states with these decay modes, but they may present
accessible targets for the continuation of LHC Run~2. Mixings between the \VL\ and SM fermions might also provide interesting
signatures in flavour and precision electroweak physics, although these are absent in the limit of small heavy-light mixing that we
consider in this paper.

Another scenario that we have considered briefly in this paper is that the lightest \VL\ fermion might provide the cosmological
cold dark matter. This is certainly possible in our Model 4, with perturbative couplings and dark matter mass in the 1-2.5 TeV range.
However, this is not the only possibility, since one is free to postulate supersymmetric versions of the \VL\
fermion scenarios considered here, in which the lightest supersymmetric particle could provide the dark matter. Indeed,
one could argue that supersymmetry could be useful to stabilize the mass of the $X$ boson
and the scale of whatever scalar field is responsible for the masses of \VL\ fermions.

\vspace{0.5cm}
{\bf Note added}

Several other papers~\cite{ambulance} on the possible $X(750)$~GeV particle appeared on the day we submitted this paper to the arXiv, some of which treat similar aspects of its interpretation.

\section*{Acknowledgements}

SARE thanks Zhengkang Zhang and Yue Zhao for useful discussions, and VS thanks Ciaran Williams for conversations on the MCFM.
The work of JE was supported partly by the London Centre for Terauniverse Studies (LCTS), using funding from the European Research Council via the Advanced Investigator Grant 26732, and partly by the STFC Grant ST/L000326/1. The work of SARE was supported partly by the DOE Grant DE-SC0007859.
The work of JQ was supported by the STFC Grant ST/L000326/1. The work of VS was supported partly by the STFC Grant ST/J000477/1.
The work of TY was supported by a Junior Research Fellowship from Gonville and Caius College, Cambridge.

\appendix

\section{Vector-Like Models in 2-Component Notation}

In this Appendix we write out explicitly the Lagrangians for Models 1-4 in two-component notation, for additional clarity about the models we consider.

\vspace{0.5cm}
\textbf{Model 1:}

In Model 1 we add a \VL\ top partner SU(2)$_L$ singlet only. The Lagrangian in both four- and two-component notation is then
\begin{align}
\mathcal{L} &= -\lambda_{STT} S \bar{T} T -\lambda_{PTT} P \bar{T} \gamma^5 T- M_T \bar{T} T - (\lambda_{HtT} \tilde{H} \bar{t}_L T + \mu_{tT}\overline{t}_{R,SM} T+ h.c.) \\ \nonumber &+ \mathcal{L}_{gauge~int.} + \mathcal{L}_{kinetic} \\
&= - (\lambda_{STT}S + M_T)\left((T_R)^{\alpha}(T_L)_{\alpha} + (T_L)^\dagger_{\dot{\alpha}}(T_R)^{\dagger \dot{\alpha}}\right) \\ \nonumber &-\lambda_{PTT}a \left( -(T_R)^{\alpha}(T_L)_{\alpha} + (T_L)^\dagger_{\dot{\alpha}}(T_R)^{\dagger \dot{\alpha}}\right)\\
&\nonumber - \left( \lambda_{HtT} \tilde{H} \left((t_L)^\dagger_{\dot{\alpha}}(T_R)^{\dagger \dot{\alpha}}\right) + \mu_{tT}\left( (t_R)^\alpha (T_L)_\alpha\right) + h.c. \right) + \mathcal{L}_{gauge~int.} + \mathcal{L}_{kinetic} \, ,
\end{align}
We list the bilinear SM-\VL\ mass mixing terms $\mu_{sm,VL}$ for completeness, but note that they can be rotated away by an appropriate choice of fields and a redefinition of Yukawa couplings.

\vspace{0.5cm}
\textbf{Model 2:}

In Model 2 we add a \VLQ\ SU(2)$_L$ doublet only. The Lagrangian in both four- and two-component notation is then
\begin{align}
\mathcal{L} &= -\lambda_{SQQ} S \bar{Q} Q -\lambda_{PQQ} P \bar{Q} \gamma^5 Q - M_Q \bar{Q} Q \\&\nonumber- (\lambda_{Qt} \tilde{H} \bar{U} t_R + \lambda_{Qb} H \bar{D} b_R + \mu_{Qq}\bar{U} t_L+\mu_{Qq}\bar{D} b_L + h.c.) \\&\nonumber+ \mathcal{L}_{gauge~int.} + \mathcal{L}_{kinetic} \\
&= - (\lambda_{SQQ}S + M_Q)\left((U_R)^{\alpha}(U_L)_{\alpha} + (U_L)^\dagger_{\dot{\alpha}}(U_R)^{\dagger \dot{\alpha}} + (D_R)^{\alpha}(D_L)_{\alpha} + (D_L)^\dagger_{\dot{\alpha}}(D_R)^{\dagger \dot{\alpha}}\right) \\ \nonumber&-\lambda_{PQQ}a \left( -(U_R)^{\alpha}(U_L)_{\alpha} + (U_L)^\dagger_{\dot{\alpha}}(U_R)^{\dagger \dot{\alpha}} - (D_R)^{\alpha}(D_L)_{\alpha} + (D_L)^\dagger_{\dot{\alpha}}(D_R)^{\dagger \dot{\alpha}}\right)\\
&\nonumber - \Bigg( \lambda_{Qt} \tilde{H} \left((U_L)^\dagger_{\dot{\alpha}}(t_R)^{\dagger \dot{\alpha}}\right)  + \lambda_{Qb} H \left((D_L)^\dagger_{\dot{\alpha}}(b_R)^{\dagger \dot{\alpha}}\right) \\ \nonumber &+ \mu_{Qt}\left( (U_R)^\alpha (t_R)_\alpha\right) + \mu_{Qb}\left( (D_R)^\alpha (b_R)_\alpha\right) + h.c. \Bigg) \\
&\nonumber + \mathcal{L}_{gauge~int.} + \mathcal{L}_{kinetic} \, .
\end{align}
Again we list the bilinear SM-\VL\ mass mixing terms $\mu_{VL,sm}$ for completeness, but note that they can be rotated away by an appropriate choice of fields and a redefinition of Yukawa couplings. 

\vspace{0.5cm}
\textbf{Model 3:}

In Model 3 we consider a combination of Models 1 and 2, with both the top partner SU(2)$_L$ singlet and the quark partner SU(2)$_L$ doublet, as well as an additional bottom partner SU(2)$_L$ singlet. Thus this model corresponds to adding $N_{QF}$ full SM-like \VLQ\ families. Bilinear mass terms mixing SM with \VL\ fields of the form $\mu_{VL,sm}$ (\VL\ doublet-SM singlet) and $\mu_{sm,VL}$ (\VL\ singlet-SM doublet) exist in principle, but can be rotated away as discussed in the text. Therefore,
we do not write them again in the Lagrangians for Models 3 and 4.

\begin{align}
\mathcal{L} &= -\lambda_{SQQ} S \bar{Q} Q -\lambda_{PQQ} P \bar{Q} \gamma^5 Q -\lambda_{STT} S \bar{T} T -\lambda_{PTT} P \bar{T} \gamma^5 T \\ \nonumber &-\lambda_{SBB} S \bar{B} B -\lambda_{PBB} P \bar{B} \gamma^5 B- M_Q \bar{Q} Q - M_T \bar{T} T - M_B \bar{B} B \\ \nonumber &-(\lambda_{QT}\tilde{H}\bar{U} T +\lambda_{QB}H\bar{D} B + h.c.) \\&\nonumber- (\lambda_{Qt} \tilde{H} \bar{U} t_R + \lambda_{Qb} H \bar{D} b_R + \lambda_{tT} \tilde{H} \bar{t}_L T + \lambda_{bB} H \bar{b}_L B + h.c.) \\&\nonumber+ \mathcal{L}_{gauge~int.} + \mathcal{L}_{kinetic} \\
& = - \lambda_{STT}S\Big((U_R)^{\alpha}(U_L)_{\alpha} + (U_L)^\dagger_{\dot{\alpha}}(U_R)^{\dagger \dot{\alpha}} + (D_R)^{\alpha}(D_L)_{\alpha} + (D_L)^\dagger_{\dot{\alpha}}(D_R)^{\dagger \dot{\alpha}} \\&\nonumber~~~~~~~~~~~~~~~~+ (T_R)^{\alpha}(T_L)_{\alpha} + (T_L)^\dagger_{\dot{\alpha}}(T_R)^{\dagger \dot{\alpha}} + (B_R)^{\alpha}(B_L)_{\alpha} + (B_L)^\dagger_{\dot{\alpha}}(B_R)^{\dagger \dot{\alpha}}\Big) \\
\nonumber &-\lambda_{PQQ}a \left( -(U_R)^{\alpha}(U_L)_{\alpha} + (U_L)^\dagger_{\dot{\alpha}}(U_R)^{\dagger \dot{\alpha}} - (D_R)^{\alpha}(D_L)_{\alpha} + (D_L)^\dagger_{\dot{\alpha}}(D_R)^{\dagger \dot{\alpha}}\right) \\ 
\nonumber&-\lambda_{PTT}a \left( -(T_R)^{\alpha}(T_L)_{\alpha} + (T_L)^\dagger_{\dot{\alpha}}(T_R)^{\dagger \dot{\alpha}}\right)-\lambda_{PBB}a \left( -(B_R)^{\alpha}(B_L)_{\alpha} + (B_L)^\dagger_{\dot{\alpha}}(B_R)^{\dagger \dot{\alpha}}\right)
\\ \nonumber &+M_Q\left( (U_R)^{\alpha}(U_L)_{\alpha} + (U_L)^\dagger_{\dot{\alpha}}(U_R)^{\dagger \dot{\alpha}} + (D_R)^{\alpha}(D_L)_{\alpha} + (D_L)^\dagger_{\dot{\alpha}}(D_R)^{\dagger \dot{\alpha}}\right) \\ \nonumber&- M_T\left( (T_R)^{\alpha}(T_L)_{\alpha} + (T_L)^\dagger_{\dot{\alpha}}(T_R)^{\dagger \dot{\alpha}} \right) - M_B \left( (B_R)^{\alpha}(B_L)_{\alpha} + (B_L)^\dagger_{\dot{\alpha}}(B_R)^{\dagger \dot{\alpha}}\right) \\ \nonumber &-\left( \lambda_{QT} \tilde{H} \left( (U_R)^\alpha (T_L)_\alpha + (U_L)^{\dagger \dot{\alpha}} (T_R)_{\dagger \dot{\alpha}}\right) + \lambda_{QB} H \left( (D_R)^\alpha (B_L)_\alpha + (D_L)^{\dagger \dot{\alpha}} (B_R)_{\dagger \dot{\alpha}}\right) + h.c.\right) \\
&\nonumber - \Bigg( \lambda_{Qt} \tilde{H} \left((U_L)^\dagger_{\dot{\alpha}}(t_R)^{\dagger \dot{\alpha}}\right)  + \lambda_{Qb} H \left((D_L)^\dagger_{\dot{\alpha}}(b_R)^{\dagger \dot{\alpha}}\right)  \\ \nonumber &+ \lambda_{tT} \tilde{H} \left((t_L)^\dagger_{\dot{\alpha}}(T_R)^{\dagger \dot{\alpha}}\right) + \lambda_{bB} H \left((b_L)^\dagger_{\dot{\alpha}}(B_R)^{\dagger \dot{\alpha}}\right) + h.c. \Bigg) \\
&\nonumber + \mathcal{L}_{gauge~int.} + \mathcal{L}_{kinetic} \, .
\end{align}

\vspace{0.5cm}
\textbf{Model 4:}

In this model we start from the particle content of Model 3, and add a full complement of SM-like vector-like leptons, including a neutral singlet \VL\ partner $N,~\bar{N}$. This model can be interpreted as postulating a \VL\ pair of $\mathbf{16}+\overline{\mathbf{16}}$ in the language of SO(10).

The lagrangian in four-component notation is
\begin{align}
\mathcal{L}&=\mathcal{L}_{Model~3} - \lambda_{SLL} S \bar{L}L - \lambda_{SEE} S \bar{E}E  -\lambda_{SNN} S \bar{N}N  \\ \nonumber &- \lambda_{PLL} P \bar{L}\gamma^5L - \lambda_{PEE} P \bar{E}\gamma^5 E  - \lambda_{PNN} P \bar{N}\gamma^5 N  \\ \nonumber&-M_{L}\bar{L}L - M_{E}\bar{E}E - M_N\bar{N}N - (\lambda_{LE} H \bar{L}^2E + \lambda_{LN} \tilde{H} \bar{L}^1N +h.c.) \\ \nonumber& - (\lambda_{Le}H \bar{L}^2 e_R + \lambda_{\ell E}H\bar{e}_L E + \lambda_{\ell N}\tilde{H} \bar{\nu}_L N +h.c.)\\ \nonumber& + \mathcal{L}_{gauge~int.} + \mathcal{L}_{kinetic} \, ,
\end{align}
which can be written in two-component notation as
\begin{align}
\mathcal{L}&=\mathcal{L}_{Model~3} - \lambda_{SLL} S\left( (L^2_R)^\alpha(L^2_L)_\alpha + (L^2_L)^\dagger_{\dot{\alpha}}(L^2_R)^{\dagger \dot{\alpha}}\right) \\ \nonumber &- \lambda_{SEE} S \left( (E_R)^\alpha(E_L)_\alpha+(E_L)^\dagger_{\dot{\alpha}}(E_R)^{\dagger \dot{\alpha}}\right)  \\ \nonumber &- \lambda_{SNN} S \left( (N_R)^\alpha(N_L)_\alpha+(N_L)^\dagger_{\dot{\alpha}}(N_R)^{\dagger \dot{\alpha}}\right)  \\ \nonumber&- \lambda_{PLL} P \left( -(L^2_R)^\alpha(L^2_L)_\alpha + (L^2_L)^\dagger_{\dot{\alpha}}(L^2_R)^{\dagger \dot{\alpha}}\right) - \lambda_{PEE} P \left(-(E_R)^\alpha(E_L)_\alpha+(E_L)^\dagger_{\dot{\alpha}}(E_R)^{\dagger \dot{\alpha}}\right) \\ \nonumber &- \lambda_{PNN} P \left( -(N_R)^\alpha(N_L)_\alpha+(N_L)^\dagger_{\dot{\alpha}}(N_R)^{\dagger \dot{\alpha}}\right)  \\ \nonumber&-M_{L}\left( (L^1_R)^\alpha(L^1_L)_\alpha + (L^1_L)^\dagger_{\dot{\alpha}}(L^1_R)^{\dagger \dot{\alpha}}+ (L^2_R)^\alpha(L^2_L)_\alpha + (L^2_L)^\dagger_{\dot{\alpha}}(L^2_R)^{\dagger \dot{\alpha}}\right) \\ \nonumber&- M_{E}\left( (E_R)^\alpha(E_L)_\alpha+(E_L)^\dagger_{\dot{\alpha}}(E_R)^{\dagger \dot{\alpha}}\right) - M_N\left( (N_R)^\alpha(N_L)_\alpha+(N_L)^\dagger_{\dot{\alpha}}(N_R)^{\dagger \dot{\alpha}}\right) \\ \nonumber &-\left( \lambda_{LE} H \left( (L^2_R)^\alpha (E_L)_\alpha + (L^2_L)^{\dagger \dot{\alpha}} (E_R)_{\dagger \dot{\alpha}}\right) + \lambda_{LN} \tilde{H} \left( (L^1_R)^\alpha (N_L)_\alpha + (L^1_L)^{\dagger \dot{\alpha}} (N_R)_{\dagger \dot{\alpha}}\right) + h.c.\right) \\ \nonumber& - \left(\lambda_{Le}H \left( (L^2_L)^\dagger_{\dot{\alpha}}(e_R)^{\dagger\dot{\alpha}}\right) + \lambda_{\ell E}H\left( (e_L)^\dagger_{\dot{\alpha}} (E_R)^{\dagger \dot{\alpha}}\right) + \lambda_{\ell N}\tilde{H} \left( (\nu_L)^\dagger_{\dot{\alpha}} (N_R)^{\dagger \dot{\alpha}}\right) +h.c. \right)\\ \nonumber& + \mathcal{L}_{gauge~int.} + \mathcal{L}_{kinetic} \, .
\end{align}
The couplings of the neutral \VL\ partner $N$ to the $S$ and $P$ fields have been written down because, despite not being relevant for the decay of $S$/$P$, they are important for the calculation of the relic density if the lightest neutral particle is stable.

 \providecommand{\href}[2]{#2}\begingroup\raggedright
\end{document}